\pdfoutput=1

\documentclass[10pt]{article}

\usepackage{listings}    
\usepackage{graphicx}    
\usepackage{subcaption}  
\usepackage{algorithm}   
\usepackage{booktabs}    
\usepackage{hyperref}    
\usepackage[htt]{hyphenat} 
\usepackage{microtype} 
\DisableLigatures{}    

\usepackage{caption}
\usepackage{setspace}
\usepackage{multirow}
\usepackage{enumerate}
\usepackage{pdflscape}
\usepackage{pgfplots} 

\usepackage{xcolor}
\definecolor{codegreen}{rgb}{0.25,0.5,0.35}
\definecolor{codegray}{rgb}{0.5,0.5,0.5}
\definecolor{codepurple}{rgb}{0.6,0,0}
\definecolor{backcolour}{rgb}{0.95,0.95,0.92}
\definecolor{colorstring}{rgb}{0.5,0,0.35}
\definecolor{rltred}{rgb}{0.5,0,0}
\definecolor{rltgreen}{rgb}{0,0.5,0}
\definecolor{rltblue}{rgb}{0,0,0.5}
\definecolor{DarkGreen}{rgb}{0.00,0.60,0.00}
\definecolor{ScarletRed}{rgb}{0.80,0.00,0.00}
\definecolor{blizzardblue}{rgb}{0.67, 0.9, 0.93}
\definecolor{green-yellow}{rgb}{0.68, 1.0, 0.18}
\definecolor{dkgreen}{rgb}{0,0.6,0}
\definecolor{gray}{rgb}{0.5,0.5,0.5}
\definecolor{mauve}{rgb}{0.58,0,0.82}
\definecolor{lightgrey}{rgb}{0.90,0.90,0.90}
\definecolor{grey}{gray}{0.75}
\definecolor{light-gray}{gray}{0.80}

\definecolor{ForestGreen}{RGB}{34,139,34}
\definecolor{asparagus}{rgb}{0.53, 0.66, 0.42}

\lstdefinestyle{mystyle}{
    escapechar=©, 
	backgroundcolor=\color{backcolour},
    basicstyle=\footnotesize\ttfamily,
   	identifierstyle=\footnotesize\ttfamily,
	commentstyle=\color{codegreen},
	keywordstyle=\color{colorstring}\bfseries,
	numberstyle=\ttfamily\color{codegray},
	stringstyle=\ttfamily\color{DarkGreen},
	breakatwhitespace=false,
	breaklines=true,
	captionpos=b,
	keepspaces=true,
	numbers=left, 
	numbersep=2pt,
	showspaces=false,
	showstringspaces=false,
	showtabs=false,
	tabsize=2
}
\usepackage{xspace}
\newcommand{\evo}{{\sc EvoMaster}\xspace}
\newcommand{\etal}{{\emph{et al.}}\xspace}

\usepackage{boxedminipage}
\newenvironment{result}%
{\smallskip
	\noindent
	\let\emph=\textbf
	\begin{boxedminipage}{\columnwidth}\begin{center}\em}%
		{\end{center}\end{boxedminipage}%
}

\usepackage{ifthen}
\newboolean{showcomments}
\setboolean{showcomments}{true} 

\ifthenelse{\boolean{showcomments}}{
	\newcommand{\nbc}[3]{
		{\colorbox{#3}{\bfseries\sffamily\scriptsize\textcolor{white}{#1}}}
		{\textcolor{#3}{\sf\small$\langle$\textit{#2}$\rangle$}}}
	
}{
	\newcommand{\nbc}[3]{}

}

\usepackage{authblk}                
\usepackage[square,numbers]{natbib} 
\usepackage{amsmath}                

\usepackage{geometry}
\usepackage{colortbl}

\usepackage{amsfonts}

\usepackage[algo2e,ruled,vlined,linesnumbered]{algorithm2e}

\newcommand{\csA}{{\textit{CS1 2021}}\xspace}
\newcommand{\csB}{{\textit{CS2 2021}}\xspace}
\newcommand{\csC}{{\textit{CS1 2023}}\xspace}
\newcommand{\csD}{{\textit{CS2 2023}}\xspace}
\newcommand{\csE}{{\textit{CS3 2023}}\xspace}

\newcommand{\nq}{}

\newcommand{\cqa}{{How difficult was it for you to set up \evo (QA1), resolve dependencies (QA2), write driver (QA3), run \evo (QA4), and execute tests(QA5)?}\xspace}

\newcommand{\cqaTimeN}{{How much time did it take you to set up \evo (QA6), resolve dependencies (QA7), write driver (QA8), run \evo (QA9), and execute tests(QA10)?}\xspace}

\newcommand{\cprA}{{How familiar are you with the web service used in the study?}\xspace}
\newcommand{\cprB}{{What is the most important feature that you would like to have in the generated	test cases?}\xspace}
\newcommand{\cprC}{{What properties are important for you to rate readability? (such as size of test cases, name of tests, grouped tests into different files, etc.) }\xspace}
\newcommand{\cprD}{{What properties are important for you to rate quality? (such as code coverage, automatically test the web services regarding input validation, scenarios (e.g., invalid inputs, business logic) to be tested automatically, etc.)}\xspace}
\newcommand{\cprE}{{Before answering the following questions, how much time did it take you to read and understand the tests generated by EvoMaster?}\xspace}
\newcommand{\cprF}{{Did you have as much time as you needed to inspect all the test cases in order to answer these questions?}\xspace}

\newcommand{\prA}{{\emph{PR\nq1: \cprA}}\xspace}
\newcommand{\prB}{{\emph{PR\nq2: \cprB}}\xspace}
\newcommand{\prC}{{\emph{PR\nq3: \cprC}}\xspace}
\newcommand{\prD}{{\emph{PR\nq4: \cprD}}\xspace}
\newcommand{\prE}{{\emph{PR\nq5: \cprE}}\xspace}
\newcommand{\prF}{{\emph{PR\nq6: \cprF}}\xspace}

\newcommand{\cqbA}{How would you like to rate the readability of the generated tests?\xspace}
\newcommand{\cqbB}{How would you like to rate the quality of the generated tests?\xspace}
\newcommand{\cqbC}{How can the generated tests be improved?\xspace}
\newcommand{\cqbD}{Describe what you like better about manually written tests than generated tests?\xspace}
\newcommand{\cqbE}{Would you keep the generated system-level tests?\xspace}
\newcommand{\cqbF}{If yes, how would you like to keep and use them?\xspace}
\newcommand{\cqbG}{{If you are using \evo, for how long (eg, minutes, hours) do you typically run it for generating test cases? (Default \texttt{--maxTime} option is 1 minute)}\xspace}
\newcommand{\cqbH}{{Based on results achieved by \evo, which option of time-cost do you prefer based on the trade-off with the achieved code coverage?\xspace}}

\newcommand{\qbA}{{\textit{QB1: \cqbA}}\xspace}
\newcommand{\qbB}{{\textit{QB2: \cqbB}}\xspace}
\newcommand{\qbC}{{\textit{QB3: \cqbC}}\xspace}
\newcommand{\qbD}{{\textit{QB4: \cqbD}}\xspace}
\newcommand{\qbE}{{\textit{QB5: \cqbE}}\xspace}
\newcommand{\qbF}{{\textit{QB6: \cqbF}}\xspace}
\newcommand{\qbG}{{\textit{QB\nq7: \cqbG}}\xspace}
\newcommand{\qbH}{{\textit{QB\nq8: \cqbH}}\xspace}

\newcommand{\cqcA}{What are the major barriers from your point of view in adopting the \evo tool?\xspace}
\newcommand{\cqcB}{Given your current infrastructure setup, how would you like to have automated system-level test generation framework integrated?\xspace}

\newcommand{\cqdA}{What is one of your current major issues/time consuming activity with manual testing that you would like to have automation for?\xspace}
\newcommand{\cqdB}{What kinds of faults are harder to detect in the system?\xspace}
\newcommand{\cqdC}{What are the most important challenges that you meet in testing?\xspace}

\newcommand{\qdA}{{\textit{QD1: \cqdA}}\xspace}
\newcommand{\qdB}{{\textit{QD2: \cqdB}}\xspace}
\newcommand{\qdC}{{\textit{QD3: \cqdC}}\xspace}

\newcommand{\vfirst}{v1.3.0\xspace}
\newcommand{\vsecond}{v1.6.1\xspace}
\newcommand{\vthird}{v2.0.0\xspace}

\newcommand{\futuredirection}{\textbf{Future direction}}

\usepackage{boxedminipage}
{\smallskip
	\noindent
	\let\emph=\textbf
	\begin{boxedminipage}{\columnwidth}\textbf{Summary}\\ \em}%
	{\end{boxedminipage}%
}

\newcommand{\sep}{,\xspace}

\title{
Fuzzing Microservices: A Series of User Studies in Industry on Industrial Systems with EvoMaster
}

\author[1]{Man Zhang}
\author[2]{Andrea Arcuri}

\author[3]{Yonggang Li}
\author[3]{Yang Liu}
\author[3]{Kaiming Xue}
\author[3]{Zhao Wang}
\author[3]{Jian Huo}
\author[3]{Weiwei Huang}

\affil[1]{Beihang University, Beijing, China}
\affil[2]{Kristiania University College and OsloMet, Norway}
\affil[3]{Meituan, Beijing, China}

\date{}

\begin{document}

\maketitle

\begin{abstract}
With several microservice architectures comprising of thousands of web services in total, used to serve
630 million customers,
companies like Meituan face several challenges in the verification and validation of their software.
The use of automated techniques, especially advanced AI-based ones, could bring significant benefits here.
\evo is an open-source test case generation tool for web services, that exploits the latest advances in the field of Search-Based Software Testing research.
This paper reports on our experience of integrating the \evo tool in the testing processes at Meituan over almost 2 years (i.e., between October 2021 and July 2023).
Two user studies were carried out in 2021 (with two industrial APIs) and in 2023 (with three industrial APIs) to evaluate
two versions of \evo (i.e., v1.3.0 and v1.6.1), respectively, in tackling the test generation for
industrial web services which are parts of a large e-commerce microservice system.
The two user studies involve in total 321,131 lines of code from these five APIs and 27 industrial participants at Meituan.
Questionnaires and interviews were carried out in both user studies with the engineers and managers at Meituan.
The two user studies demonstrate clear advantages of \evo (in terms of code coverage and fault detection) and the urgent need to have such a fuzzer in industrial microservices testing.
To study how these results could generalize, a follow up user study was done in 2024 (with \evo \vthird) with five engineers in the five different companies.
Our results show that, besides their clear usefulness, there are still many critical challenges that the research community needs to investigate to improve performance further.

\end{abstract}


\section{Introduction}

In microservice architectures~\cite{newman2021building}, large enterprise systems are split in hundreds/thousands of connected web services.
This software architecture is widely applied in industry, to enable rapid and frequent delivery of the services in production.
Considering the large number of services and their complex interactions, there are several challenges in the verification and validation of such systems.
Automated testing techniques (such as \evo~\cite{arcuri2018evomaster,arcuri2021evomaster,zenodo200evomaster}) could be one possible manner to address such challenges.

EvoMaster~\cite{arcuri2018evomaster,arcuri2021evomaster,zenodo200evomaster} is an open-source tool for \emph{fuzzing} Web APIs at the system testing level  with \emph{search-based} techniques.
 It has been evaluated as the most performing tool in recent studies on REST API fuzzers~\cite{Kim2022Rest,zhang2023open}.
With the term ``fuzzing''~\cite{zeller2019fuzzing,zhu2022fuzzSuvery,godefroid2020fuzzing} we simply mean the \emph{ability of generating system level test cases that can detect faults} (e.g., program crashes).
For generating test cases efficiently, we rely on the output of several decades of scientific research in the field of \emph{Evolutionary Computation}, in particular in the context of \emph{Search-Based Software Testing} (SBST)~\cite{HaJ01,harman2012search,mcminn2011search}.
Furthermore, to be actually of use for practitioners, test cases should be \emph{readable} (i.e., easy to understand), as one of their main goals is to help \emph{debugging} when these tests detect faults.
Note: in the scientific community there is no consensus on how these terms Fuzzing and SBST are and should be used~\cite{guizzo2023fuzzing}; it is an ongoing debate.
According to the aforementioned definition, we state that \evo is a \emph{search-based fuzzer}.

When experimenting with test case generation techniques, one challenge is that large enterprise systems are rarely present in open-source repositories.
The performance (such as line coverage and fault finding) of \evo has been mainly assessed on open-source APIs~\cite{arcuri2019restful,arcuri2019sql,arcuri2021tt,zhang2021adaptive,zhang2021resource,zhang2023open} (such as APIs collected in the GitHub repository EMB~\cite{EMB,icst2023emb}).
But, most of these open-source APIs are web services that work in isolation, and not part of a microservice architecture~\cite{EMB,icst2023emb}.

As part of industry-driven research~\cite{garousi2019characterizing}, \evo has been evaluated on industrial APIs as well (e.g., in~\cite{arcuri2021enhancing}).
But those were part of small systems.
Microservices architecture is now widely applied in industry, but  microservices also pose a set of unique challenges in their testing.
One of industrial partners, i.e.,
Meituan\footnote{https://about.meituan.com/en}, a large e-commerce company seeks an automated API testing solution for optimizing their development/testing processes for microservices.
As researchers,
to better address this industrial problem, it is important to identify the limitations of our current approach (i.e., \evo) and challenges that industry faces  that the research community can address further.
For a test generation tool, it is not only a matter of finding faults.
How actual practitioners use those tools, and how they integrate them in their development processes, is of paramount importance~\cite{nourry2023human}.
This is particularly the case for large enterprise systems, where several non-technical challenges are present as well (e.g., communication issues when there are hundreds of engineers, and dependencies among different services
developed by different teams).
Therefore, 
we carried out two user studies with our industrial partner (i.e., Meituan), involving real industrial system under tests (SUTs) which are parts of a large-scale e-commerce system, and involving employees who can provide feedback from the point of view of practitioners through questionnaires and interviews.
Our main objectives in this study are:
\begin{itemize}
	\item to evaluate the \emph{usability}\footnote{the extent to which a new approach (such as \evo) is easy to apply and use in industrial practice} of \evo in industrial contexts;
	\item to assess the \emph{effectiveness}\footnote{the extent to which objectives can be achieved by the approach (such as \evo) in testing large-scale industrial applications} of \evo at tackling test generation problems in a real, large industrial setting;
	\item to investigate essential features for enabling its further adoption into industrial practice;
	\item further, to study existing testing challenges in industry which could be addressed by the research community.
\end{itemize}
The first user study was conducted at the starting phase of the collaboration with Meituan in 2021. 
In the first user study,  
we employed \evo on two real industrial services, performed a detailed review of results achieved by \evo (e.g., its generated test cases), and conducted a questionnaire and interviews with eight industrial practitioners at Meituan.
The questions in our interviews/questionnaire 
in the user study are a superset of an existing study in the unit testing domain~\cite{seip2017industrial}, which enable us to do some forms of meta-analysis.
Regarding feedback on the 
usability
of the tool, for the engineers at Meituan, there did not seem to exist any major difficulty in applying the tool on their industrial applications.
For effectiveness, results on the two chosen APIs show that \evo achieved 
up to 33.5\% line coverage, and on average (i.e., arithmetic mean over two APIs
) 26 faults could be detected by manually reviewing the generated tests.
Such results demonstrated the potential benefits achieved by our approach to our industrial partner and can further strength the research-industry collaboration. 

The second user study was a follow-up study conducted at Meituan in 2023 with our latest technique (i.e., addressed the most important challenges for Meituan by designing a novel approach for enabling a native fuzzing of APIs developed with Remote Procedure Call (RPC)~\cite{zhang2023rpc}) for demonstrating how the identified challenges have been addressed and what current progress of integrating our fuzzer (academic prototype) into industrial practice is after 1.5 years (i.e., from October 2021 to May 2023).
In the follow up study in 2023, we further included some new important questions that were not considered in the first study.
Compared to the results obtained in the first study, readability of generated tests has been improved due to the support for RPC, such as native function invocation and thrown exceptions (used to identify potential faults) shown in the tests, and tests grouped by distinguishing whether they represent potential faults.
Regarding the effectiveness on the three RPC APIs, \evo achieved up to 33.4\% line coverage and identified on average (i.e., arithmetic mean) 64.8 potential faults automatically.
With the further feedback from the participants, they pointed out that
readability should be further improved by having meaningful strategies to group and name tests, and the code coverage also needs to be improved by testing more combinations of various RPC functions and interfaces in order to better cover the business logic of their systems.

Besides assessing the usability and effectiveness of \evo, with these two user studies, we also investigated the potential adoption of our approach to industrial practice through the research-industry collaboration.
We report on the major updates which our industrial partner and us have developed for enabling integration of \evo into their industrial development processes.
\evo  has now been integrated into the industrial pipelines of our industrial partner, and is used daily for testing tasks impacting hundreds of engineers at Meituan.
Moreover, we discuss lessons learned in terms of testing setup, testing criteria, locating faults, 
and assertion generation while conducting this study.
Furthermore, we summarize important common challenges we faced and future directions in the fuzzing of industrial microservices.

One limitation of these two user studies is that they are done in industry, with only one enterprise, i.e., Meituan.
In contrast to novel algorithm advances (which require no evaluation by practitioners in industry to provide a scientific contribution), or user studies with students (which can provide generalizable results~\cite{salman2015students,falessi2018empirical}), user studies in industry with software engineering practitioners must be carried out in several different enterprises to have scientific value.
To address this threat to validity, a third user study outside of Meituan was carried out in 2024.
This involved five practitioners from five different Chinese enterprises, using online questionnaire, and it employed the same questions as the second user study from 2023.
As we do not have formal collaboration agreements with these five companies, the five participants volunteered to take their free time to be part of this user study.
Results showed that responses to \emph{usability} were much more negative than those we received at Meituan.
The task of writing drivers to enable white-box testing appeared to be significantly more difficult for participants outside of Meituan.
In terms of \emph{effectiveness}, all participants stated that they did not have enough time to inspect the generated tests in details.
This indicates the challenge of conducting this type of user study in companies where there is no formal collaboration, and so the participants cannot spend too much of their free time on such a study (they could spend a few hours, but not more).
In terms of \emph{integration} and \emph{existing challenges}, as they are independent of \evo, we found that most of the responses from outside of Meituan are consistent with what collected in the first two user studies conducted at Meituan.

The main contribution of this paper is an empirical study of the application of an academic fuzzer in a real industrial setting, targeting the domain of large scale microservice architectures, including how software engineers would integrate such fuzzers in their development processes.
%
The study reports on our experience of two user studies conducted with our industrial partner Meituan in assessing the
usability and effectiveness of our academic fuzzer in an industrial setting, investigating the adoption of our fuzzer in industrial practices, and identifying important challenges existing in this industrial context.
A third study was carried out outside of our industrial partner aimed at generalizing our finding to other enterprises.

We believe that the evaluation of scientific research in actual industrial settings, albeit in a single company, is an essential step toward addressing the significant gap between research and practice.
However, we do acknowledge that this is a very controversial statement in the Software Engineering Research community.
On the one hand, research work only evaluated in the lab might rely on assumptions that might not hold true in practice.
On the other hand, industry-relevant research with collaborations with industry partners take a massive amount of time~\cite{davis2023s}, and it seems quite rare that this kind of work involve more than
one industrial partner at a time~\cite{ko2015practical}.
 This  can lead to less generalizable results.
Both kinds of studies are important, but, considering the current trends in Software Engineering research,
\emph{``Generalizability is overrated''}~\cite{briand2017case}.
There is a large body of scientific work in designing novel fuzzing techniques, but their application by actual practitioners in industry is a software engineering aspect that is not widely investigated~\cite{nourry2023human}.

The paper is organized as follows.
We introduce background information in Section~\ref{sec:back}, and discuss related work in Section~\ref{sec:related}.
Section~\ref{sec:design} presents our design of the empirical study followed by experimental results discussion in Section~\ref{sec:results}.
Lessons learned and common challenges are summarized in Section~\ref{sec:lessons}.
We clarify threats to validity in Section~\ref{sec:threats} and conclude the paper in Section~\ref{sec:conclude}.

\section{Background}\label{sec:back}

	\subsection{Web APIs}

	Application Programming Interface (API) is a common practice in building enterprise microservices.
	For example, REST and RPC are two methods used widely for facilitating communications (typically based on HTTP)  among services in an enterprise~\cite{newman2021building}.

\subsubsection{REST and OpenAPI}

	REpresentational State Transfer (REST) is a set of  architectural guidelines rather than a protocol~\cite{fielding2000architectural}.
	 It is composed of a collection of design principles used to develop modern web services, using the HTTP/HTTPS protocol.
	For instance, to better manage resources in web services, REST suggests that resources should be identified by the Uniform Resource Identifiers (URIs), and implementations of operations on resources should always follow semantics of HTTP methods, e.g., GET should be used to retrieve resource(s). 
	Nowadays, REST has been applied in many companies, such as Google\footnote{https://developers.google.com/drive/v2/reference/}, Amazon\footnote{http://docs.aws.amazon.com/AmazonS3/latest/API/Welcome.html}, and Twitter\footnote{https://dev.twitter.com/rest/public}.
	
	OpenAPI\footnote{https://www.openapis.org/} (previously known as Swagger) is a specification that standardizes descriptions of REST APIs.
	Such a specification allows both human and machine to understand how the API works and how to interact with its services. 
	An example as below shows how an endpoint of a REST API can be described using OpenAPI.
	In this example, information of a pet can be retrieved by an HTTP GET request with a given pet ID of type \texttt{integer}, and the ``responses'' also lists possible HTTP status codes, e.g., ``404'' status code indicates that a pet with the given ID cannot be found.
\begin{figure}

\begin{lstlisting}
/pet/{petId}": {
	"get": {
		"summary": "Find pet by ID",
		"description": "Returns a single pet",
		"operationId": "getPetById",
		"produces": [
			"application/json",
			"application/xml"
		],
		"parameters": [
			{
				"name": "petId",
				"in": "path",
				"description": "ID of pet to return",
				"required": true,
				"type": "integer",
				"format": "int64"
			}
		],
		"responses": {
			"200": {
				"description": "successful operation",
				"schema": {"$ref": "#/definitions/Pet"}
			},
			"400": {
				"description": "Invalid ID supplied"
			},
			"404": {
				"description": "Pet not found"
			}
		}
	},
	"delete": {...
\end{lstlisting}
\caption{An example of OpenAPI specification}
\label{fig:openapi}
\end{figure}
\subsubsection{RPC}

Remote procedure call (RPC) is a communication protocol which is widely applied in enterprise services (such as Netflix, Alibaba), especially for distributed systems.
An RPC API is defined by one or more RPC interfaces.
Note that the implementation of the business logic is often organized by such interfaces.
Each interface is further composed of a set of functions which are exposed to the other services.
Unlike REST, there does not exist amy widely accepted standards for describing RPC APIs, such as OpenAPI for REST.
There exists diverse frameworks, such as Dubbo\footnote{https://dubbo.apache.org} by Alibaba, gRPC\footnote{https://grpc.io} by Google, 
Tars\footnote{https://tarscloud.org/} by Tencent, 
Thrift\footnote{https://thrift.apache.org/} by Apache, and companies might also develop their own in-house RPC frameworks for satisfying their needs, e.g.,
Meituan developed MThrift,
a modified version of Apache Thrift.

\begin{figure}
	\centering
	\includegraphics[width=0.6\linewidth]{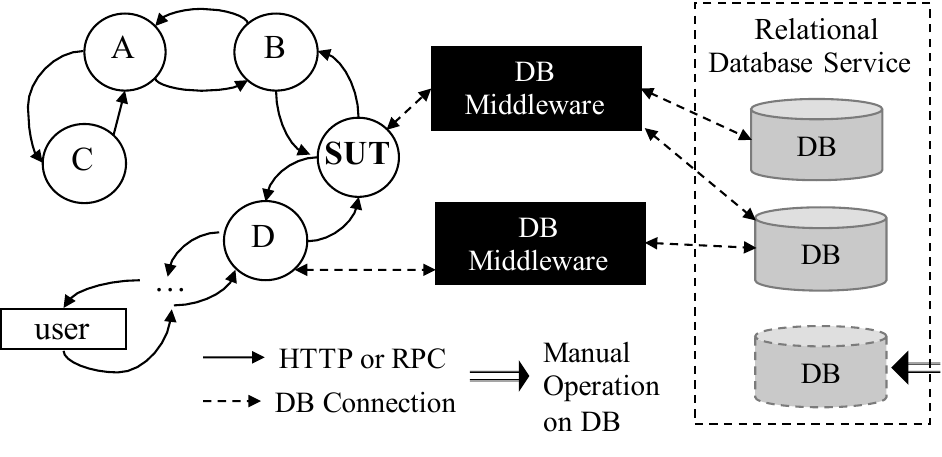}
	\caption{An example of the SUT with its connected services}
	\label{fig:sut}
\end{figure}

Figure~\ref{fig:sut} shows an example of a large microservice architecture with APIs, such as RPC and REST.
With RPC communication, a stub is needed for supporting interactions between services.
For instance, if the service \emph{SUT} needs to access a function \texttt{foo} provided in \emph{B} (Figure~\ref{fig:sut}), the stub of \emph{B} is instantiated in the \emph{SUT}, that then invokes it as \texttt{result=stubB.foo()}.
In this paper,  the SUTs from Meituan are all RPC Java web services interacting with many other services and databases (as \textit{SUT} shown in Figure~\ref{fig:sut}).
The SUTs were chosen by the team at Meituan we are in contact with, as those services implement their core business.

\subsection{EvoMaster}

\label{sec:evomaster}

\begin{figure}
	\centering
	\includegraphics[width=0.6\linewidth]{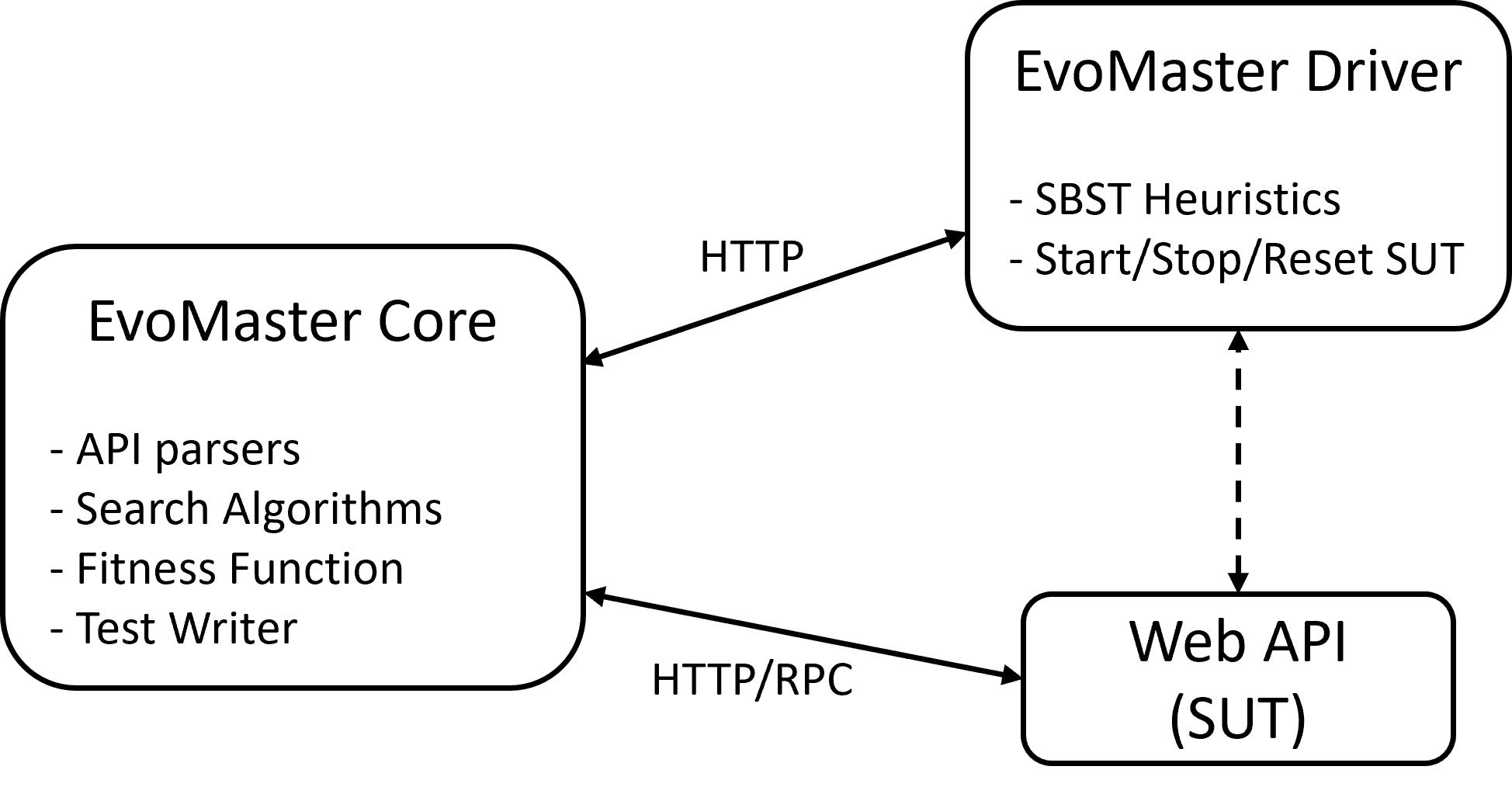}
	\caption{An overview of \evo}
	\label{fig:evomaster}
\end{figure}

\evo is an open-source testing tool designed for addressing system-level test case generation for enterprise web services (currently supports REST, GraphQL APIs and RPC APIs)~\cite{arcuri2021evomaster,belhadi2022graphql,zhang2023rpc}.
An overview of \evo is shown in Figure~\ref{fig:evomaster}.
It is a search-based tool, using evolutionary algorithms to generate effective test cases.
The tool enables both white-box~\cite{arcuri2017restful,arcuri2018test,arcuri2019restful,js2022,zhang2023javascript} (for APIs running on the JVM and NodeJS) and black-box testing~\cite{arcuri2020blackbox}.

\evo is composed of two parts (see Figure~\ref{fig:evomaster}), i.e., \emph{Driver} and \emph{Core}.
\textit{Driver} (only used when performing white-box testing) implements  bytecode instrumentation, which enables the collection at runtime of code coverage and advanced white-box heuristics.
To manipulate SUTs, \emph{Driver} also defines interfaces which requires to be extended manually by users in order to configure how to start/stop/reset the SUTs.
\textit{Core} provides a set of search algorithms (such as MIO~\cite{arcuri2018test} and MOSA~\cite{dynamosa2017}) to generate tests, defines fitness functions to guide the search, and extracts info on how to access the SUT (e.g., OpenAPI~\cite{Swagger} for REST APIs).

\subsubsection{The MIO Algorithm}

The Many Independent Objective (MIO)~\cite{arcuri2018test} algorithm is specialized for system-level test case generation in the context of white-box testing.
It is the default search algorithm for white-box \evo.
Let us briefly summarize it here in this section.
The algorithm is inspired by (1+1) EA~\cite{LeY07}, which only employs sampling and mutation operators.
Pseudo-code of the algorithm is shown in Algorithm~\ref{algo:mio}.

In the MIO algorithm, an \emph{individual} is a test case, and the test case is composed of a sequence of actions (e.g., HTTP actions for REST API) to test the SUT.
To guide the testing,
we defined \textit{fitness function} $\delta$ with \textit{testing targets}, which comprise all lines, branches, fault finding and domain specific targets (such as status code for REST).
Each of them is a target to be optimized (i.e., covered) during the evolutionary search.
For instance,  for \texttt{x == 42}, we define two branch targets for the condition, i.e., one is for \texttt{true}, and other is for \texttt{false}.
If the execution reaches the statement, we consider the two branch targets \emph{Reached}.
The \texttt{true} branch target can be considered as \emph{Covered} only if the condition is evaluated as \texttt{true},
otherwise the other branch \texttt{false}  is considered as \emph{Covered}.
Covering both branches would require the statement to be executed at least twice.

In the REST domain, 500 status code in the response could used to represent a potential fault~\cite{zhang2023open,Kim2022Rest}, and the set of valid status codes for each endpoint is specified in the schema.
Thus, each 500 status combined with distinct last executed line in the SUT is identified as a distinct potential fault.
Any status code present in the response represents that a status code target is \emph{Covered}, and an occurrence of invalid status code further considers that a mismatched schema fault is identified (i.e., \emph{Covered}).

Each target has a population for it with a maximum size (denoted as $n$).
The MIO algorithm starts with an empty population (denoted as $S$).
At each iteration, the algorithm would either sample a new individual or mutate an existing individual in the evolving populations, controlled by a probability $P_r$.
If the individual reaches a new target, the MIO algorithm would create a new population for it (e.g., a new population $T_k$ for the newly covered target $k$).
If a target is covered with the individual, the size of its corresponding population would be reduced to 1, and the target would not be considered further in the search.
In the context of web service testing, the individual is a test which is a sequence of HTTP calls.
After a certain percentage of search budget is used (denoted as $F$), the MIO algorithm uses a \textit{focused search},  where it increases the probability to mutate the individuals which have recent improvement, instead of exploring other areas of the search landscape.
At the end of the search, the MIO algorithm outputs a test suite for the SUT, comprising of the best evolved test cases for each testing target (denoted as $A$).

\begin{algorithm2e}
	\DontPrintSemicolon
	\SetAlgoLined
	\SetKwInOut{Input}{Input}\SetKwInOut{Output}{Output}
	\Input{Stopping condition $C$, Fitness function $\delta$, Population size $n$, Probability for random sampling $P_{r}$, Start of focused search $F$}
	\Output{Archive of optimized individuals $A$}
	\BlankLine
	$S \leftarrow SetOfEmptyPopulations(\, )\,$ \;
	$A \leftarrow \{\}$ \;
	\While{$\neg$$C$}{
		\eIf{$P_{r} > rand(\,)\,$}{
			$p \leftarrow RandomIndividual(\,)\,$\;
		}{
			$p \leftarrow SampleIndividual(S)\,$\;
			$p \leftarrow Mutate(p)\,$\;
		}
		\ForEach{element $k$ $\in$ ReachedTargets($p$)}{
			\If{NewTarget(k)}{
				$S \leftarrow S \cup T_{k}$\;
			}
			$T_{k} \leftarrow T_{k}\cup \{p\}$\;
			\uIf{$IsTargetCovered(k)\,$}{
				$UpdateArchive(A, p)\,$\;
				$S\leftarrow S \setminus T_{k}$\;
			}
			\ElseIf{$|T_{k}| > n$}{
				$RemoveWorst(T_{k}, \delta)\,$\;
			}
		}
		$UpdateParameters(F, P_{r}, n)\,$
	}

	\caption{Pseudo-code of the MIO Algorithm~\cite{arcuri2018test}}\label{algo:mio}
\end{algorithm2e}

\subsubsection{Integrated Novel Techniques}

Throughout the years, since its inception in 2016, \evo has been extended with various novel techniques~\cite{arcuri2020sql,arcuri2021tt,zhang2021resource,zhang2021adaptive,belhadi2022graphql,arcuri2024advanced} to serve a more comprehensive testing in the context of white-box testing for enterprise systems.

Enterprise web systems typically interact with databases and employ SQL to specify operations on it.
To test such systems, states of interacted databases (such as what data are in the databases) would possibly have a direct impact on code coverage.
\evo was enhanced with a SQL handling technique~\cite{arcuri2020sql}, which defines SQL query heuristics, and enables SQL execution monitoring and direct data  insertion into the database with SQL commands from the tests.
With such technique, \evo is capable of producing more effective tests with generated SQL commands based on the SQL heuristics to directly manipulate states of the SUT.
In this context, a test (also referred as an individual) with additional SQL commands could result in a long and complex chromosome to evolve.
To have an effective mutation for handling such individuals,  an \textit{adaptive weight-based hypermutation}~\cite{zhang2021adaptive} was developed that comprises a set of novel strategies to adaptively manipulate the number of genes to mutate, select genes to mutate, and guide how to mutate the values in these genes.

In the context of white-box testing, the \textit{flag problem} is a common issue, i.e., lack of guidance to perform the search due to boolean expressions.
To address this issue, \textit{testability transformation}~\cite{arcuri2021tt,arcuri2024advanced} techniques have been developed in \evo that enable to transform the source code of the SUT (with \textit{replacement} methods), for providing better gradient values, e.g., for \textit{branch distance} computations~\cite{Kor90}.
In addition, test inputs are tracked in the transformed methods that further provides additional info (referred as  taint analysis) to better generate test data.
Moreover, \evo uses specific transformations particularly for  handling web service testing, e.g., to handle cases in which the OpenAPI schemas are underspecified.

REST is one of the most applied techniques for building web services in industry.
To enable an automated testing for it, \evo has been integrated with a set of techniques based on REST domain.
REST test generation problem was reformulated as a search problem~\cite{arcuri2019restful} by defining various types of genes corresponding to REST schema, i.e., OpenAPI.
In addition, the fitness function for MIO was enhanced with additional considerations on HTTP status codes and fault detection (i.e., 500 status code).
Moreover, to better sample tests for REST, a smart sampling technique was developed with pre-defined structures.
Furthermore, by further exploiting REST domain knowledge, a \textit{resource-dependency based} MIO~\cite{zhang2021resource} was proposed that is composed of resource-based sampling, mutation and dependency heuristics for producing tests with effective resource handling.
Besides REST APIs, \evo also enables the fuzzing of GraphQL APIs~\cite{belhadi2022graphql} (in 2022) and RPC APIs~\cite{zhang2023rpc} (in 2023).

\section{Related Work}
\label{sec:related}


Most of recent approaches for automated web service testing are developed for REST APIs in the context of black-box testing~\cite{golmohammadi2023testing}, e.g.,~\cite{viglianisi2020resttestgen,restlerICSE2019,Godefroid2020Restlerdata,karlsson2020QuickREST,laranjeiro2021black,martinLopez2021Restest,wu2022icse}.
For REST APIs, OpenAPI~\cite{Swagger} provides a well-structured and machine-readable schema language to specify how to access these web services.
Such schemas provide a base to enable automated testing techniques to access the system under test (SUT).
For example,
RESTler proposed by Atlidakis \etal~\cite{restlerICSE2019}
can automatically produce a sequence of requests to test REST APIs.
The sequence is decided based on either a statistical analysis on the schema with request types, or a dynamic analyses on feedback received from previously executed requests.
In addition, RESTler is now integrated with different testing techniques for regression testing~\cite{godefroid2020differential}, security aspect testing~\cite{atlidakis2020checking}, and test data generation~\cite{Godefroid2020Restlerdata}.
Viglianisi \etal~\cite{viglianisi2020resttestgen} developed
RESTTESTGEN that employs a dependency analysis on the schema to better generate test orders and inputs.
The approach comprises two components, i.e., nominal tester and error tester, for finding defects by testing the SUT against nominal scenarios and error scenarios.
Laranjeiro \etal~\cite{laranjeiro2021black} proposed bBOXRT for robustness testing in the context of black-box testing by testing the REST APIs with various invalid inputs. The invalid inputs are generated based on a set of mutation rules on data type with the schema.
RESTest~\cite{martinLopez2021Restest} developed by Martin-Lopez \etal is a black-box testing framework which has been integrated with different techniques such as fuzzing test inputs, adaptive random testing and constraint-based testing to generate tests for REST APIs.
The constraint-based testing is performed based on inter-parameter dependencies analysis on the schema.
RestCT~\cite{wu2022icse} proposed by Wu~\etal is a black-box combinatorial testing approach for REST APIs.
The approach consists of two phases for generating orders and inputs of requests with OpenAPI specifications.
Hatfield-Dodds and Dygalo developed Schemathesis~\cite{hatfield2022deriving} using property-based testing techniques.
The tool derives structure and semantics of APIs for enabling REST API testing.

Considering that a black-box testing approach is independent from the programming language and source code, there are  available more industrial SUTs to be used for evaluating it, compared with a white-box testing approach.
For example, live services on the internet can be used for these kinds of experiments.
Many industrial services (such as Microsoft Azure, Office365 cloud services, Google Drive, Spotify), and hundreds of real REST APIs listed on public API aggregators, have been used to conduct experiments when evaluating the aforementioned black-box testing techniques.
Several real faults were found, e.g., based on the returned HTTP status codes, in a process that is also known as \emph{fuzzing}.

Testing REST APIs share some similar challenges with testing RPC APIs, e.g., related to the handling of interactions with databases and external services.
But,
at the time of our first experiment in 2021, there was no existing fuzzer for RPC APIs.
Extending \evo to support RPC APIs~\cite{zhang2023rpc} was a direct follow up from that first study, to fill such important gap in the research literature.
Since that time, a work on supporting the testing of RPC APIs has been published in a workshop in 2023~\cite{veldkamp2023grammar}, based on a master thesis published in 2022~\cite{veldkamp2022msc}.
Such work presents a black-box fuzzer based on OpenRPC specifications~\cite{OpenRPC}, which is not applicable to the systems of Meituan (as not using OpenRPC).

With recent studies~\cite{Kim2022Rest,zhang2023open} conducted for studying existing fuzzers (including the fuzzers described above and \evo), white-box \evo achieved the best performance on open-source APIs.
In addition,
to the best of our knowledge, there does not exist in the literature any industrial evaluation  for white-box system testing of microservices.
This paper fills this important gap in the research literature.

Regarding industrial evaluations, there exist some recent work to conduct empirical studies on automated unit testing in industrial settings related to financial applications~\cite{seip2017industrial} and embedded systems~\cite{zhang2018smartunit}.
Regarding system-level testing, the tool evaluations have been mainly investigated for user interface (UI) testing on ERP applications
~\cite{brunetto2021introducing} and mobile applications~\cite{sapienz2018,wang2018empirical}.
Other testing techniques, such as mutation testing~\cite{beller2021would,petrovic2018state}, fault debugging~\cite{zhou2018fault} and flaky tests~\cite{Godefroid2019Flaky} have also been evaluated in industry.
However, none of such work provides the same type of contribution we give in this paper:
an empirical study in which engineers and testers in industry apply a test generation tool on their industrial systems by themselves (without any intervention from researchers, using the publicly available documentation of the tool), and then evaluate the quality (e.g., readability) and benefits (e.g., code coverage and fault detection) of the obtained results (i.e., generated test cases) on their industrial systems, involving tens/hundreds of thousands of lines of code for the business logic.


There has been work on evaluating the readability of automatically generated tests.
This has been the case for unit testing, with tools such as EvoSuite~\cite{daka2015modeling}.
However, these studies usually involve only students (e.g., 30 students in~\cite{daka2015modeling}).
As the subjects of these studies are single classes, which have a limited number of lines of code, this type of experiments on \emph{unit testing} are feasible.
Similarly, since 2023 the Java Test Case Generation Tool Competition\footnote{https://sbft23.github.io/tools/java}
has started to use readability to score the competing tools.
This was done by hiring external people (possibly engineers), which are not the authors of the tested classes.
However, to the best of our knowledge, there is no work in the scientific literature that has studied the readability of automatically generated \emph{system level tests}, in particular not when applied on large industrial systems.

	In this article, we report on 3 user studies over 3 years, involving a total of 32 practitioners.
	In the software engineering research literature, there have been several studies involving human subjects.
	An early survey~\cite{sjoberg2005survey} considering the years from 1993 to 2002, based on a selection of 5~453 articles from flagship software engineering research venues, shows that only 103 (i.e., less than 2\%) studies included human subjects.
	Afterwards, Ko \etal~\cite{ko2015practical} analyzed 1 701 articles published from 2001 to 2011 in flagship software engineering research venues.
	They showed that 345 of these articles (i.e., 20\%) contain tool evaluations with human participants.
	When considering the time period from 2011 to 2018, based on a survey~\cite{siegmund2020mastering} of 1 584 articles published in flagship venues, the number of articles involving studies with human subjects was 397 (i.e., 25\%).
	There is a clear increase of this type of studies throughout the years.

	Some examples of this type of study include the work of  Briand \etal~\cite{briand2014traceability}, where they presented a user study on traceability involving 20 students.
	Scanniello \etal~\cite{scanniello2015documenting} reported on 4 controlled empirical studies (from 2011 to 2013), involving 88 participants.
	Of those, 25 were professionals in industry, whereas the others were students.
	The study involved \emph{``comprehend a nontrivial chunk of an open-source software system''}~\cite{scanniello2015documenting}.
	Fraser \etal~\cite{fraser2015does} investigated the utility of unit test generators with 2 studies involving 92 students and 5 industrial developers on 4 Java classes from open-source projects.
	Paulweber \etal~\cite{paulweber2021specifying} carried out a study on object-oriented abstractions with 98 students.
	Santos \etal~\cite{santos2021family} conducted a family of studies (8 with students and 4 in industry) on TDD for a total of 411 participants.
	Five ``toy tasks''~\cite{santos2021family} were used for the experiments.

	In the review from 1993 to 2002~\cite{sjoberg2005survey}, 72\% of the analyzed articles exclusively involved students.
	Such percentage decreased to 23\% in the 2001 to 2011 review~\cite{ko2015practical} (where 56\% included at least 1 professional, and the remaining did not provide details on the participants).
	Whether students or professionals in industry should be used for this type of study has been long debated.
	For example, to shed light on this issue,  Salman \etal~\cite{salman2015students} ran an empirical study on TDD with 17 students and 24 professional on ``toy tasks''~\cite{salman2015students}.
	In their experiments, \emph{``neither of the subject groups performs better than the other when they apply a new technology during experimentation''}~\cite{salman2015students}.
	Falessi \etal~\cite{falessi2018empirical} stated \emph{``Using students as participants remains a valid simplification of reality needed in laboratory contexts''}.
	They carried out a survey among ``experts'' to study how the role of students and professionals in studies with human subjects in software engineering research is seen.
	Using students instead of professionals can simplify the complexity of running this kind of study.
	As such, they are more common among researchers.
	As stated by Vegas \etal
	\emph{``experiments in industry are thin on the ground. Of the few existing cases, most
		are 1-1 (running one experiment at one company), just a few are
		n-1 (running n experiments at one company) and still fewer are 1-
		n (running one and the same experiment at n companies)''}~\cite{vegas2015difficulties}.
	They reported on the difficulty of running experiments related to TDD in 3 different companies, including
	\emph{``Professionals were troublesome, undermotivated, and performed worse than students''}~\cite{vegas2015difficulties}.
	However, although experiments were carried out in industry, the software subjects were green-field exercises (and not the industrial systems developed at those 3 companies).

	Carrying out studies with human subjects is complex and time consuming.
	This can put off researchers from conducting this kind of research.
	For example, as stated by Ko \etal: \emph{``Recent research has also shown that many software engineering researchers view this form of tool evaluation as too risky and too difficult to conduct, as they might ultimately lead to inconclusive or negative results''}~\cite{ko2015practical}.
	To address this issue, Davis \etal~\cite{davis2023s} carried out interviews with 26 researchers to define a taxonomy of 18 barriers researchers encounter in this kind of study.
	They then proposed 23 solution strategies to address 8 of those barriers.
	
	What presented in this paper is significantly different from what reported in this section so far.
	None of the aforementioned work deal with \emph{user studies with professionals in industry on tasks involving their industrial systems}.
	Some exceptions exist, like the study on Copilot made by the researchers at GitHub~\cite{ziegler2024measuring}.
	However, without a systematic review of the literature (e.g., the $103+345+397=845$ articles identified in~\cite{sjoberg2005survey,ko2015practical,siegmund2020mastering}) we cannot say for sure how rare such kind of study is.

\section{Experiment Design}
\label{sec:design}

\subsection{Research Questions}

To investigate \evo in industrial settings and understand industrial challenges in the testing of web services, 
we conducted an empirical study to answer the following research questions:

\begin{description}

	\item[{\bf RQ1}:] How difficult is to set up a tool like \evo in real industrial settings?
		\begin{description}
			\item[{\bf RQ1.1}:] How difficult do industrial practitioners think is to use \evo?
			\item[{\bf RQ1.2}:] What challenges do exist in setting up \evo in industrial settings?
		\end{description}
	\item[{\bf RQ2}:] How does \evo perform in real industrial settings?
	\begin{description}
		\item[{\bf RQ2.1}:] How does \evo perform on industrial SUTs in terms of line coverage and fault detection?
		\item[{\bf RQ2.2}:] How effective do industrial practitioners consider the automatically generated tests?
	\end{description}
	
	\item[{\bf RQ3}:] How is \evo integrated into industrial development process?
	\begin{description}
		\item[{\bf RQ3.1}:] What are major barriers of adopting \evo from the viewpoint of industrial practitioners?
		\item[{\bf RQ3.2}:] 
		What is the collaborative process of integrating \evo into industrial pipeline at our industrial partner?
		\item[{\bf RQ3.3}:] How would  industrial practitioners  like to use \evo?
	\end{description}
	
	\item[{\bf RQ4}:] What are the most important testing challenges that practitioners meet in their testing context?
	

%

\end{description}


\subsection{Experimental Tasks}
\label{subsec:tasks}

Our RQs are designed to investigate \evo in industrial contexts in terms of \emph{usability} (RQ1), \emph{effectiveness} (RQ2), \emph{integration} (RQ3) and \emph{existing challenges} that industry currently faces (RQ4).
Table~\ref{tab:exp-design} outlines the tasks for answering each RQ.
We carried out three user studies with industrial practitioners.
Two of these studies (i.e., User Study 2021 and User Study 2023) were conducted in collaboration with our industrial partner, Meituan.
The third user study (i.e., User Study 2024) involved practitioners from other five companies with whom we do not have existing collaborations, aiming to generalize findings obtained at Meituan.

Regarding tasks, as show in Table~\ref{tab:exp-design}, assessments of \emph{usability} (RQ1), \emph{effectiveness} (RQ2), \emph{integration} (RQ3) of \evo in industrial contexts were designed to consider perspectives of both researchers (RQ1.2, RQ2.1 and RQ3.2) and practitioners (RQ1.1, RQ2.2, RQ3.1 and RQ3.3).
Benefiting from research-industry collaboration, as authors of academic prototype (i.e., researchers), we gain access to valuable information from industrial SUTs and have the opportunity for direct communication.
More specifically, as the authors of \evo, we can check enterprise APIs chosen by our industrial partner and the quality of drivers to ensure that \evo was properly used for performing white-box testing on the enterprise APIs. 
With this opportunity, we shared our experience for answering RQ1.2.
In addition, with information about the SUTs and direct feedback from industrial partner, it allows us to better analyze the line coverage and faults detected by \evo on industrial SUTs.
This enables us to better understand \evo's performance in real-world settings, as outlined in RQ2.1.
Without such a collaboration, it would be difficult to perform such analyses.
Therefore, this kind of analysis was excluded in User Study 2024.

Note that, in the first attempt to use \evo at Meituan, i.e., in User Study 2021, one employee involved in the task for RQ2.1 (see Table~\ref{tab:exp-design}) together with the authors of \evo did manually review the tests generated by \evo and their coverage reports.
This was done in order to identify the potential benefits \evo could offer.
Such a review was also necessary to assess \evo's fault detection capabilities to avoid issues of \evo on identifying faults of RPC APIs using domain knowledge of REST in User Study 2021.
To minimize the efforts required from the employee, we selected one test suite for each SUT, opting for the test suite that achieved the best result for review.
Moreover, this collaboration also enables us to discuss our perspectives in the collaborative process of integrating \evo into the industrial development process, addressing RQ3.2.
Regarding assessments conducted from practitioners (RQ1.1, RQ2.2, RQ3.1 and RQ3.3), the analyses were performed based on responses of practitioners in questionnaires/interviews.

RQ4 (\emph{existing challenges}) aims to study the challenges that practitioners face.
It was conducted only from the practitioners' perspective with questionnaires and interviews.
Identified challenges can be further addressed in \evo and may also be useful for researchers in this field to better  understand industrial problems.

\begin{table}
	\centering
	\caption{Experimental tasks for each RQ}
	\label{tab:exp-design}
	\small
	\begin{tabular}{p{0.04\linewidth} p{0.55\linewidth}  |p{0.06\linewidth} p{0.12\linewidth} | p{0.1\linewidth}}
		\cline{3-5}
		& & \multicolumn{2}{l|}{User Study 2021} & User Study\\ 
		& & \multicolumn{2}{l|}{User Study 2023} & 2024\\ 
		\toprule
		RQs  & Tasks & Resear -chers & Practitioners (Meituan) & Practitioners (Others) \\ \hline
		
		\textit{RQ1.1} &  Ask participants to follow the \evo documentation to fuzz services they are familiar with; then, collect their feedback based on their experience using \evo  & $\times$ &  \checkmark & \checkmark\\ \cline{2-5}
		\textit{RQ1.2}&  Conduct a preliminary study of \evo on enterprise APIs and report researchers' experience of making the tool feasible for use in industrial setting of \textbf{industrial partner} &  \checkmark & $\times$ & $\times$ \\ \hline
		\textit{RQ2.1} & Analyze line coverage and fault detection achieved by \evo on industrial SUTs \textbf{from industrial partner}  &  \checkmark & \checkmark  (2021) \phantom{Use} $\times$ (2023) & $\times$ \\ \cline{2-5}
		\textit{RQ2.2}&  Collect participants' opinions on tests generated by \evo on industrial SUTs  &  $\times$ & \checkmark &  \checkmark \\\hline
		\textit{RQ3.1} &  Collect feedback for adopting \evo from participants' viewpoint & $\times$  &  \checkmark  (2021) \phantom{Use} $\times$ (2023) & \checkmark\\ \cline{2-5}
		\textit{RQ3.2} &  Report researchers' experience of integrating \evo into \textbf{industrial partner}'s development process & \checkmark  &  $\times$ & $\times$ \\ \cline{2-5}
		\textit{RQ3.3}&  Collect participants' preference for applying \evo in industrial development process &  $\times$ & \checkmark &  \checkmark \\\hline
		\textit{RQ4} & Collect participants' feedback on existing challenges they face  & $\times$ & \checkmark & \checkmark \\
		\bottomrule
		
	\end{tabular}
	
\end{table}

\subsection{User Studies at Meituan}

	The two user studies at Meituan were performed at the starting point of the collaboration (i.e., User Study 2021) and at a time point when we addressed an important industrial problem specific to Meituan (i.e., User Study 2023), respectively.
	In the user studies at Meituan, our industrial partner managed the studies and decided which SUTs and participants to involve.

\subsubsection{SUTs}
\label{subsec:subject}

Meituan Select is a large-scale e-commerce platform for community group bulk buying (e.g., fresh vegetables, meat) that is a part of Meituan, and its products have covered 2~600 cities and counties in China.
Products can be bought directly from farmers or distributors, and then be delivered to the community.
With this e-commerce platform, its daily order volume is more than 30 millions.
More than 630 million users are registered and use this e-commerce platform.
The platform is constructed with hundreds of web services (referred as microservices) for, e.g., ordering, allocation, packing, delivery, and payments.

Table~\ref{tab:sut-info} presents detailed descriptive 
information of all five SUTs employed in the two user studies for studying effectiveness of \evo on enterprise APIs at Meituan, e.g., the number of Java class files, lines of codes and the number of exposed endpoints (i.e., methods that can be invoked via RPC).
In 2021, we firstly conducted the user study with two of those services at Meituan, denoted as \csA and \csB.
In 2023, the second  study was conducted with three cases studies, denoted as \csC, \csD and \csE.
Note that \emph{CS1} and \emph{CS2} refer to the same services. 
But, in 2023, we could not conduct the experiment with the SUTs using the same version as 2021, as these services (not only the two services under test but also all the  interacted services and databases) have been significantly evolved since then to satisfy Meituan's business requirements (see Table~\ref{tab:sut-info}).
%
%
\#Services is the number of other services that the SUT directly interacts with (as \emph{B} and \emph{D} in Figure~\ref{fig:sut}).
Among them, \#U  is the number of its direct upstream services which the SUT depends on (e.g., \emph{B}), and \#D is the number of its direct downstream services, which call and use the SUT (e.g., \emph{D}).
Note that here we only report the number of services with direct communications.
For instance, the upstream services of \emph{SUT} have further upstream services, e.g., \emph{A} for \emph{B} in Figure~\ref{fig:sut}.

Regarding the databases, in industrial settings where there is the need to scale to \emph{hundreds of millions} of customers, the applied database is often distributed.
At Meituan, this is achieved by sets of databases with an ad-hoc Relational Database Service (RDS), developed internally by Meituan to meet their scale needs.
Connections between the SUT and databases are managed by sets of database middlewares.
For instance, as shown in Figure~\ref{fig:sut}, a connection between \emph{SUT} and databases might be further distributed to multiple data sources (e.g., read, write) to different databases.
In addition, the RDS audits and monitors all SQL commands executed on the databases.
For example, in the first experiment, we found that \texttt{TRUNCATE} and \texttt{DROP} commands are forbidden from the web services, as the databases and tables are allowed to be created only manually through an audit process at Meituan.
In Table~\ref{tab:sut-info}, we also report 
the number of tables and the number of rows of data in the databases for all chosen web services.

All these web services chosen as SUTs in this paper are services whose previous versions are actively used in production at Meituan.
Before running the experiments for this paper, no known bugs and faults were present (as all discovered bugs are promptly fixed).


\begin{table}
	\centering
	\caption{Descriptive statistics of industrial SUTs at Meituan}
	\label{tab:sut-info}
	\resizebox{1\linewidth}{!}{
		\begin{tabular}{lllrrrrrr}
			\toprule
			SUT & Original & Type of &\#Scripts & File LOCs &\#Endpoints &\#Services (U, D) &\#Tables & \#RowsOfData \\
			&  Type & Fuzzing & &  & & & &  \\
			\midrule
			\csA & RPC & REST & 245 & 32,393  & 33& 18 (14, \phantom{1}4) &  142  & 256,024       \\
			\csB & RPC & REST & 98   &  12,152  & 13& 11 (\phantom{1}7, \phantom{1}4)  & 17&    1,840       \\
			\csC & RPC & RPC & 421 & 57,209  & 62 & 27 (21, \phantom{1}6) & 210   &  401,291      \\
			\csD & RPC & RPC & 185 & 26,353  & 20 &  33 (23, 10)  & 19    & 65,915     \\
			\csE & RPC & RPC & 308 & 193,024  & 63 & 24 (22, \phantom{1}2) &  8 & 1,017,504       \\
			\midrule
			\textit{Total} & &  & 1,257 &    321,131 &  190 &  113 (87, 26)&     268       &    1,742,574      \\
			\bottomrule
		\end{tabular}
	}
	\begin{spacing}{0.8}
		\raggedright \footnotesize -Services(U,D) represents a number of services that the SUT directly interacts with, \#U is a number of upstream services and \#D is a number of downstream services; \#RowsOfData is a number of rows of existing data in related tables.
	\end{spacing}
\end{table}

\subsubsection{User Study 2021}
\label{subsub:us2021}
The microservices at Meituan are built with Remote Procedure Call (RPC), a widely applied technique in industrial applications.
To the best of our knowledge, there did not exist any tool which was available to enable automated testing for RPC APIs in 2021, when we first introduced \evo at Meituan (this can be regarded as one challenge identified).
As REST API fuzzing and RPC API fuzzing share parts of domain knowledge in terms of Web APIs testing (such as database interactions) and white-box system level testing (such as code coverage and fault detection) and \evo has been evaluated as the most performing tool on open-source REST APIs~\cite{Kim2022Rest,zhang2023open}, our industrial partner considered that it was worth to apply \evo for assessing its performance in testing their services.
A temporary solution to use \evo was to build an additional layer to adapt RPC APIs to REST APIs (i.e., implement a one-to-one mapping from an RPC function to a REST endpoint).
However, implementing such a layer was straightforward, as discussed with our industrial partner.
In industry, it is not uncommon to build a ``proof-of-concept'' to evaluate the feasibility of an approach, before investing a considerable amount of resources.
	For example, for a first validation it was easier and quicker to build a simple REST API layer on top of 2 existing APIs,
	instead of
building a fuzzer for RPC APIs from scratch, as none existed in 2021.
	The former takes just a short amount of time (in the order of minutes, at most few hours), but has to be done for each single API, whereas the latter took several months, but needs to be done only once.
	Requiring to manually write REST layers is not a scalable solution, nor it is user-friendly compared to directly fuzz the RPC APIs.
	It was simply done just for experimentation sake.
Therefore, we conducted the first empirical study of \evo in industry by fuzzing RPC APIs with REST's strategies (see \emph{Type of Fuzzing} \csA and \csB in Table~\ref{tab:sut-info}) at Meituan in 2021.

\subsubsection{User Study 2023}
\label{subsub:us2023}
As the potential benefits of \evo was demonstrated in the first 
user study, we continued our research with our industrial partner and addressed  one of the most important challenges, i.e., fuzzing RPC APIs~\cite{zhang2023rpc}, in \evo \vsecond~\cite{zenodo161evomaster}.
	In such work~\cite{zhang2023rpc}, we provided the scientific contribution of designing the first white-box approach for fuzzing RPC APIs in the literature.
	Experiments were carried out on a series of APIs at Meituan, where performance and effectiveness were measured based on criteria such as code coverage and fault detection.
	However, such empirical study had no human subject.
	Its focus was on algorithmic analysis, and not on the human aspects of introducing and applying these kinds of techniques in industry among practitioners.
	A main difference here, compared to~\cite{zhang2023rpc}, is that we employed questionnaire for assessing \evo with additional point of view from the industrial partner.
	Furthermore, the authors of \evo were not involved in running the tool (the participants had to do it by themselves, based only on the publicly available documentation of the tool).


Once the significant problem (i.e., fuzzing RPC) was addressed in \evo for Meituan,
to perform a followup user study of \evo, we conducted the second user study at Meituan that applied the same questions as the study in 2021 for assessing 
usability and \textit{effectiveness} of \evo \vsecond, with three industrial APIs in 2023 (see \csC, \csD, and \csE in Table~\ref{tab:sut-info}).
	However, this second study is not a replication of the first study.
	The main difference is that in this second user study we used the novel introduced native support for RPC APIs in \evo.
	The generated tests call the RPC APIs directly.
	There was no need to create a REST layer anymore, and potential faults are identified based on the domain knowledge of RPC API rather than REST.

\subsubsection{EvoMaster Settings}
\label{subsec:cs-setting}
To investigate effectiveness of \evo in the two user studies at Meituan,
we conducted two experiments for fuzzing enterprise APIs using \evo \vfirst (discussed in Section~\ref{subsub:us2021}) and \evo \vsecond (discussed in Section~\ref{subsub:us2023}) respectively.
In the User Study 2021, as the first attempt to apply \evo, the experiment was ran on a local machine of an employee at Meituan, thus, we can only conduct it once due to such resource constraints. 
In the User Study 2023, as \evo is integrated into the testing pipelines at Meituan, we were able to repeat the second experiment multiple times, i.e., repeated 10 times.

Regarding time budget settings, as discussed with the industrial partner, testing performed on their CI pipeline typically takes a couple of minutes.
From a point of view of developers, since they
might wait for implementing other tasks, the time cost for the fuzzer is preferably less than 1 hour. 
However, as a search-based method, the specified search budget in \evo would have a major impact on its performance.
With a consideration on time constraints in industry, therefore, we set three stopping criteria (i.e., 30 minutes, 1 hour and 10 hours) in 
these experiments. 
Note: this is rather different from the typical amount of 24 hours used in fuzzing literature.
Hence, the two experiments of running \evo on the five SUTs took 368 hours, i.e., $ (0.5h + 1h + 10h) \times 2 $\emph{APIs} $ \times 1 + (0.5h + 1h + 10h) \times 3 $\emph{APIs} $ \times 10 $, which took more than 15 days.

\subsubsection{Industrial Participants}
In the two user studies, we involved eight employees at Meituan in 2021 and 19 employees at Meituan in 2023.
As a result of the collaboration, employees at Meituan were allowed to participate in these user studies during their working hours.

Table~\ref{tab:ind-participate} represents information about all these industrial participants, i.e., positions (see \emph{Position}), years of working in testing in industry (see \emph{\#Years}), and the size of group the participant leads (see \emph{\#Group}).
Note that terms in \emph{Position} are based on the job title terminology used  internally at Meituan.
The roles of all participants vary over seven different positions, which are responsible for various tasks, such as testing, development and product.
This could provide diverse viewpoints to this industrial evaluation.
More specifically,
the \emph{Director of Software Quality (DSQ)} participant is in charge of all testing departments at  Meituan Select.
He is  involved in the interview study mainly to provide viewpoints to challenges relating to their business scope.
The \emph{Principal Software Engineer} (PSE) participant is the department manger of internal testing tool development at  Meituan Select, who can share opinions from the standpoint of applicability and further integration of the fuzzer and challenges they meet in order to enable testing of industrial services.
The \emph{Quality Assurance Manager} (QAM) participants manage testing tasks and are responsible for various specific services, such as services for inventory management at warehouse and payment.
They could give us feedback on preference of applying \evo to their testing tasks, and share their  difficulties/experience in managing testing tasks for various kinds of industrial services.
The \emph{Quality Assurance Engineer} (QAE) participants were involved based on their familiarity with the selected SUTs.
They mainly handle testing tasks for specific services, such as write test cases and perform manual testing.
Thus, they could give feedback on the generated tests and share their experience in daily tasks which fuzzers like \evo aim to automate.
Besides feedback from practitioners who work in software testing, we also involved \emph{Product Manager} (PDM), \emph{Development Manager} (PDM) and \emph{Developer} (DPE) at Meituan for assessing \evo from different viewpoints.

\begin{table}
	\centering
	\caption{Description of Industrial Participants at Meituan}
	\label{tab:ind-participate}
	\small
	\resizebox{.99\linewidth}{!}{
	
	\begin{tabular}{lrrrrc}
		\toprule
		Position &  \#Par User Study 2021 &  \#Par User Study 2023 & \#Years &\#Group &Abb. \\
		\midrule
		\textit{Director of Software Quality} & 1 &   & 15+ & 200+& DSQ \\
		\textit{Principal Software Engineer}   & 1&   & 11--12 &30--50&  PSE    \\
		\textit{Quality Assurance (QA) Manager}  & 3& 3 & 8--10 &10--25&   QAM \\
		\textit{Quality Assurance (QA) Engineer}               & 3 &10 & 0--10 & & QAE\\
		\emph{Product Manager}  & & 1 & 11 & 10 & PDM\\
		\emph{Development Manager}  & & 2 & 10-15 & 10-35 & DPM \\
		\emph{Developer}  & & 3 & 2-6 & & DPE\\
		\midrule
		\emph{Total}  & 8 &  19 & & \\
		\bottomrule
	\end{tabular}
	
		}
	\begin{spacing}{0.8}
		\raggedright \footnotesize \#Par represents the number of participants in the experiments in either 2021 or 2023, \#Years represents years of working in testing in industry, and \#Group is a number of members in the group the participant leads.
	\end{spacing}
\end{table}


\subsection{User Study outside of Meituan}

	To generalize the results obtained at Meituan to other enterprises,
	we conducted the third user study in 2024 outside of Meituan (User Study 2024) with \evo \vthird.
	Note that there is no difference in terms of capability of fuzzing Web APIs between \vthird and \vsecond. 
	Since we do not have other collaborations with additional companies,
	one of the authors contacted few acquaintances in their professional network in China.
	Five responded positively, willing to take some of their time to participate in this user study.
	Then, the third study
	was carried out with five practitioners from five different Chinese enterprises by online questionnaire.
	One of enterprises is called \emph{Hebei Happy Consumer Finance Co., Ltd.} that mainly engages in financial-related business, such as offer loans to consumers.
	However, we did not obtain permission to disclose the names of the other four companies. 
	In addition, since we do not have collaboration agreements with these companies, it is not feasible to ask participants to conduct the experiment as what we did at Meituan, e.g., fuzz APIs with 10 hours' time budget.
	Thus, in User Study 2024, the participants themselves can decide which SUT to use and what time budgets to set.
	Moreover, without the collaboration agreements, we were also unable to obtain permission to provide statistics on the five employed Web APIs (three RPC APIs and two REST APIs), and assess SUTs and results.
	Therefore, we cannot report detailed statistics in this paper, and exclude the SUTs in analyzing results of \evo from researchers' perspectives (i.e., RQ1.2, RQ2.1 and RQ3.2).

	Regarding the participants, based on information they provided, the five practitioners have different job roles, i.e., 1 Development Manager (DPM), 1 Developer (DPE), 1 Software Architect (SWA), and 2 Testers/Quality Assurance Engineers (Tester/QAE).

\subsection{Question Design for Questionnaire/Interview}
\label{subsec:questionnaire}

One of the main tasks in an industrial evaluation is to collect feedback from the industry practitioners (all tasks relating to practitioners' perspectives in Table~\ref{tab:exp-design}).
In software engineering research, typical methods to collect such feedback are based on questionnaires and interviews conducted with employees at these companies~\cite{lethbridge2005studying,stol2018abc}.
In this study, to empirically assess the fuzzer \evo and evaluate its further potential integration in industrial settings,
we designed a set of questions for the questionnaire and interviews
that collect responses from industrial participants
in terms of \emph{usability} (for RQ1.1), \emph{effectiveness} (for RQ2.2), \emph{integration} (for RQ3) and \emph{existing challenges} (for RQ4) in the three use studies as shown in Table~\ref{tab:allques}.
The interviews were performed only with \emph{RQ3} and \emph{RQ4} that involved 1 DSQ, 1 PSE and 2 QAM (see \textsuperscript{\dag} in Table~\ref{tab:allques}).
%
5-Point likert scale~\cite{likert1932technique} is one of the typically methods to measure perceptions and opinions of humans in questionnaires and surveys, and it has been applied in various studies in software engineering~\cite{seip2017industrial,eck2019understanding,piantadosi2020does}.
In this study,
we employ a 5-Point likert scale~\cite{likert1932technique} to collect participants' feedback on assessing difficulty of using \evo (\emph{QA1-5}), readability (\emph{QB1}) and quality (\emph{QB2}) of tests generated by \evo, and participants' familiarity on the SUT (\emph{PR1}).
Such questions were not used in the interviews.

\begin{table}
	\caption{Questions in questionnaire and interview and number of participants for answering each question in each user study. Note that ``NA'' indicates No Answer meaning that the question is not used in the user study, and \textsuperscript{\dag} indicates that participants answered the questions through interviews.}
	\label{tab:allques}
	\small
	\resizebox{1.0\linewidth}{!}{
		 
		\begin{tabular}{p{0.11\linewidth} p{0.07\linewidth}  p{0.70\linewidth} |p{0.05\linewidth}| p{0.05\linewidth} | p{0.05\linewidth}}
			\toprule
			& \# & Questions & \multicolumn{3}{c}{\#Par User Study}\\\cline{4-6}
			& &  & 2021 & 2023 & 2024\\
			\midrule
			Usability &\emph{QA1-5} & \cqa &   1: & 10: & \\
			& & [ ] \emph{Very Difficult} [ ] \emph{Difficult} [ ]\emph{ Moderate} [ ] \emph{Easy} [ ] \emph{Very Easy} & {\footnotesize 1QAM}  & {\footnotesize 5QAE}  &  \\ \cline{3-4}
			& {\emph{QA\nq6-10}} & \cqaTimeN & NA & {\footnotesize 2QAM 2DPE 1DPM} &  \\\cline{1-5}

			Effective  &\textit{PR\nq1} & \cprA & \multirow{9}{*}{NA}& & \\
			-ness& & [ ]\emph{1-Very Unfamiliar} [ ]\emph{2-Unfamiliar} [ ]\emph{3-Moderate} [ ]\emph{4-Familiar} [ ]\emph{5-Very Familiar} & & &\\
			& \textit{PR\nq2} & \cprB & & & \\
			& & [ ] \emph{Achieved code coverage} [ ] \emph{Detected faults} [ ] \emph{Readability} [ ] \emph{Other \_\_}  & & &\\
			& \textit{PR\nq3} & \cprC & & &\\
			& \textit{PR\nq4} & \cprD & & &\\
			& {\textit{PR\nq5}} & \cprE & &15: &\\
			& {\textit{PR\nq6}} & \cprF & &{\footnotesize 8QAE 3QAM} & 5:\\
			& & {[ ] \emph{Yes} [ ] \emph{No}} & & {\footnotesize 2DPE} & {\footnotesize 1DPE} \\\cline{2-4}
			
			& \textit{QB1} & \cqbA &  & {\footnotesize 1DPM}&{\footnotesize 1DPM}\\
			& & [ ] \emph{Very Low} [ ] \emph{Low} [ ] \emph{Moderate} [ ] \emph{High} [ ] \emph{Very High} & &{\footnotesize 1PDM} &{\footnotesize 1SWA}\\
			& \textit{QB2} & \cqbB & & &{\footnotesize 2Teste}\\
			& & [ ] \emph{Very Low} [ ] \emph{Low} [ ] \emph{Moderate} [ ] \emph{High} [ ] \emph{Very High} & & &{\footnotesize r/QAE}\\
			& \textit{QB3} & \cqbC & & &\\
			& \textit{QB4} & \cqbD & 5:& &\\
			& \textit{QB5} & \cqbE &{\footnotesize 3QAE} & &\\
			& & {[ ] \emph{Yes} [ ] \emph{No}} &{\footnotesize 2QAM} & &\\
			& \textit{QB6} & \cqbF & & &\\\cline{3-4}
			& {\emph{QB\nq7}} & {\cqbG} & \multirow{3}{*}{NA} & &\\
			& {\emph{QB\nq8}} & \cqbH & & & \\
			& & {[ ] \emph{30 minutes} [ ] \emph{1 hour} [ ] \emph{10 hours}} & & & \\ \cline{1-5}

			Integration & \textit{QC1} & \cqcA 	& 8: & NA & \\\cline{5-5}
			& \textit{QC2} & \cqcB  	& {\footnotesize 3QAE 1QAM} & 19: {\footnotesize 10QAE} &\\\cline{1-3}

			Existing Challenges & \textit{QD1} & \cqdA & {\footnotesize 1\textsuperscript{\dag}DSQ 1\textsuperscript{\dag}PSE} & {\footnotesize 3QAM 3DPE} &\\
			 & \textit{QD2} & \cqdB & {\footnotesize 2\textsuperscript{\dag}QAM} & {\footnotesize 2DPM} &\\
			 & \textit{QD3} & \cqdC &  & {\footnotesize 1PDM}&\\
			\hline
			\# Answers	&  \multicolumn{5}{r}{(\emph{QAs}: $5 \times 16 + 5 \times 15$ ) $+$ (\emph{PRs}: $ 6 \times 20$ ) $+$ (\emph{QBs}: $ 6 \times 25 + 2 \times 20$ ) $+$ (\emph{QCs}: $ 13 + 32$ ) $+$ (\emph{QDs}: $ 3 \times 32 $) $=606$}\\
			\bottomrule
	\end{tabular}}
	
\end{table}

\subsubsection{Usability}
To evaluate the 
usability of \evo, we defined \emph{QA} questions according to steps of using \evo as shown in Table~\ref{tab:allques}.
On \evo website\footnote{\url{www.evomaster.org}},
documentation and examples including some training videos are provided about how to get started with \evo.
This includes the following five steps.
\begin{itemize}
	\item \emph{Set up}: there are three options to set up \evo, i.e., use a released runnable jar file, install it with installers for Windows/OSX/Linux since version 1.2.0, or build it from source code;
	\item \emph{Resolve dependencies}: it might need to resolve conflicts of dependencies (such as different versions of used libraries) between \evo and SUT;
	\item \emph{Write driver}: to use \evo in white-box mode, it requites to write a driver for extending pre-defined methods which enables starting/stopping/resetting of the SUT programmatically (see Figure~\ref{fig:evomaster});
	\item \emph{Run \evo}: start the fuzzer from command line for automatically generating tests for the SUT;
	\item \emph{Execute tests}: execute generated tests on the SUT.
\end{itemize}
Given such documentation, we asked the industrial participants 
to use the fuzzer with their services, then collect their feedback. 

In 2021, at the starting phase of collaboration with Meituan, due to limited resources, \evo was assessed by one participant who provided his/her ratings of difficulty for the five steps of using \evo (see \emph{QA1-5} in Table~\ref{tab:allques}).
In 2023, with more resources from Meituan allocated to the collaboration, the usability assessment in the User Study 2023
was conducted with 10 participants.
In addition, we defined five new questions (i.e., \emph{QA6-10} in Table~\ref{tab:allques}) to assess the time cost of using \evo according to the five steps.
In User Study 2024, all \emph{QA}s were employed.  

\subsubsection{Effectiveness}
\label{subsubsec:eff}
To evaluate effectiveness of \evo from the point of view of industrial practitioners, we carried out experiments of fuzzing enterprise APIs with \evo, and then
conducted a questionnaire about generated tests, given to industrial participants. 

The questions were designed to collect feedback on the tests regarding their readability, their quality, how they can be improved and whether they keep them as \emph{QB}s shown in Table~\ref{tab:allques}.
Given the results on the effectiveness of \evo, it is also interesting to investigate practitioners' preferences regarding time cost settings.
Then, in User Study 2023 with \evo \vsecond, we designed two further questions (i.e., \emph{QB7} and \emph{QB8}) to collect responses related to time budget that practitioners would like to use.

In the second user study, we further designed a set of pre-check questions, as \emph{PR}s shown in Table~\ref{tab:allques}.
The goal was to obtain information regarding their familiarity on the SUT they analyze (\emph{PR1}), preference to assess the generated tests (\emph{PR2--4}), time spent on performing this experiment and whether they had enough time (\emph{PR5--6}).
With the three SUTs applied in 2023, for each one, we involved five employees at Meituan (in total 15 participants for all three SUTs).
To better assess the generated tests, ideally there would need to involve more people who are familiar with the SUT to test.
However, in industry, services are typically developed by different groups, and, in quality assurance departments, they typically divide tasks among different employees bases on the services.
Then, there might not be many employees  who are familiar with the same SUT.
The five employees per SUT is the maximum that Meituan could reach, without having to involve employees that are not familiar with those APIs.

In User Study 2024, the same questions as User Study 2023 were asked to the five participants.

\subsubsection{Integration}
We defined two questions (i.e., \emph{QC1} and \emph{QC2}) in Table~\ref{tab:allques} in our questionnaire and interview for inquiring potential integration and usage of \evo in industrial settings.

\emph{QC1} relates to barriers in adopting \evo that was involved in User Study 2021 and User Study 2024.
We executed it in User Study 2023, because \evo has been integrated into the testing pipelines at Meituan.
In User Study 2024, we included \emph{QC1} to collect the feedback from practitioners outside of Meituan.
Regarding \emph{QC2}, we asked it in all of the three user studies.

\subsubsection{Existing Challenges}
The research task for RQ4 aims at understanding existing testing difficulties in industry, and discuss potential solutions that tools like \evo could help to tackle.
To achieve this goal, we defined three questions (i.e., \emph{QD1-3}) as shown in Table~\ref{tab:allques} and asked them to all industrial participants in the three user studies.

\section{Experiment Results}
\label{sec:results}

\subsection{Results for RQ1: Usability}
\label{subsec:rq1}
The goal of this RQ1 is to evaluate the 
usability of \evo in industrial settings in the view of industrial participants (i.e., RQ1.1) and discuss challenges we faced (i.e., RQ1.2).

\subsubsection{Results of RQ1.1}
\label{subsubsec:rq11}
{\textbf{Results at Meituan.}}
In the User Study 2021, one employee at Meituan (i.e., 1 QAM) applied \evo to test enterprise RPC APIs accessed with exposed REST endpoints.
His/her responses for \emph{QA1-5} on difficulty rates of using \evo (i.e., \emph{QA1-5}) are \emph{Easy} for setting up, writing driver, running the tool and executing tests, and \emph{Moderate} for resolving dependencies.

\begin{figure}
	\centering
	\begin{subfigure}{.47\textwidth}
		\includegraphics[width=\linewidth]{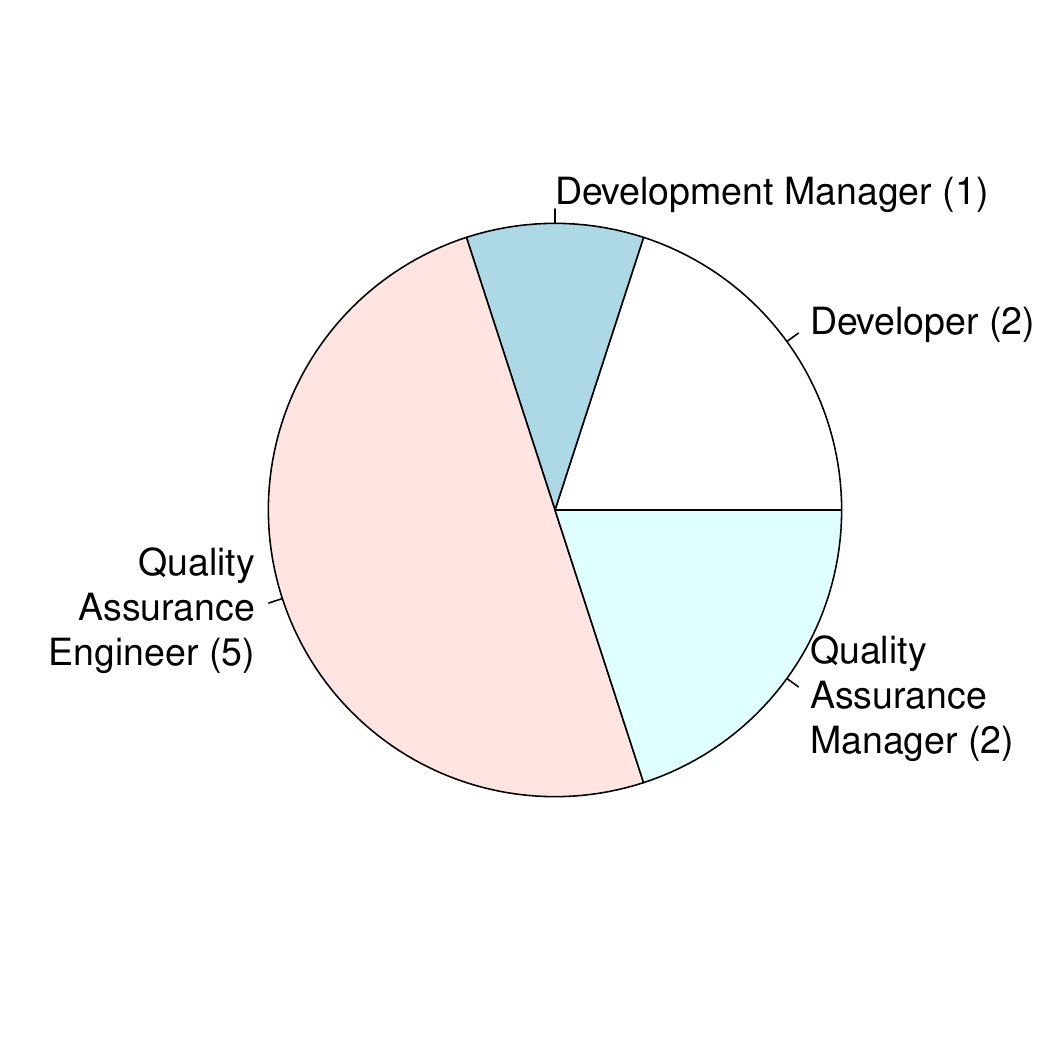}
		\vspace{-3\baselineskip}
		\caption{Participants in 2023}
		\label{subfig:appEx2023Employee}
	\end{subfigure}\hfill
	\begin{subfigure}{1\textwidth}
		\includegraphics[width=\linewidth]{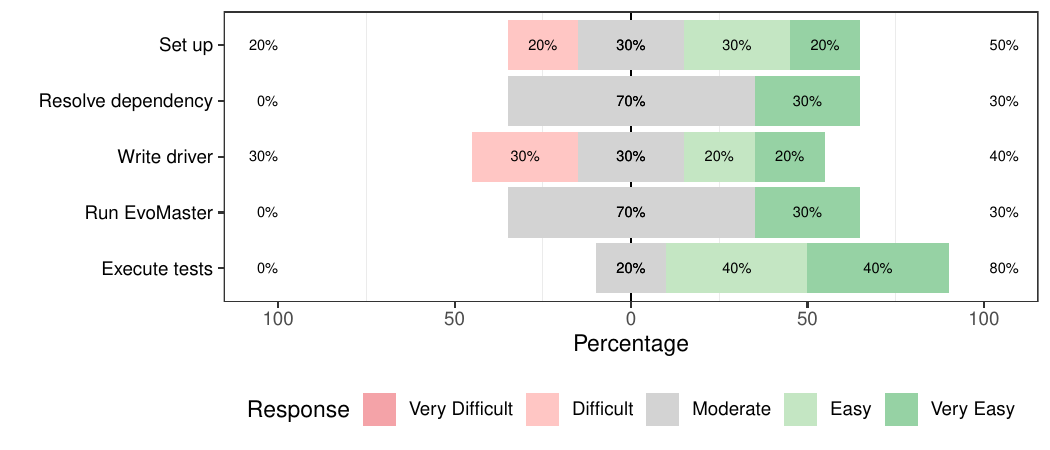}
		\caption{Diverging stacked barplot for responses of difficulty rates using 5-points likert-scale collected in 2023. Note that positive responses (i.e., ``easy'') extending to the right and negative responses (i.e., ``difficult'')  to the left from the diverging point.}
		\label{subfig:appEx2023Rates}
	\end{subfigure}\hfill
	\caption{Answers provided by industrial partners about the difficulties on applying \evo (QAs)}
	\label{fig:rq1}
\end{figure}

In 2023, as the version \vsecond supported RPC API fuzzing, we conducted the User Study 2023 for assessing its usability with a native support for fuzzing
enterprise RPC APIs.
The 
User Study 2023 involved 10 participants at Meituan, and positions of the participants are shown in Figure~\ref{subfig:appEx2023Employee}.
Results of answers as the same set of questions in 2021 (i.e., \emph{QA1-5}) are shown in Figure~\ref{subfig:appEx2023Rates}, and answers of new questions in terms of time cost the participants spent to use \evo (i.e., \emph{QA\nq6-10}) are shown in Table~\ref{tab:appExTime}.


%
%
%

\begin{table}
	\centering
	\small
	\caption{Time (minutes) spent by participants for applying \evo in 2023}
	\label{tab:appExTime}
		
		\begin{tabular}{ l rrrrr r}\\ 
\toprule 
Position & Setup & Resolve dependency & Write driver & Run EvoMaster & Execute tests & Total\\ 
\midrule 
DPE & 180 & 30 & 60 & 30 & \underline {180} & 480\\ 
DPE & 20 & \textbf{1} & \textbf{3} & 10 & 5 & \textbf{39}\\ 
DPM & 20 & 2 & 5 & 10 & \textbf{2} & \textbf{39}\\ 
QAE & 240 & 60 & 60 & 30 & 120 & 510\\ 
QAE & 240 & 60 & 60 & 30 & 120 & 510\\ 
QAE & \underline {300} & 90 & \underline {120} & \underline {90} & 120 & \underline {720}\\ 
QAE & 120 & 30 & 30 & 30 & 30 & 240\\ 
QAE & 120 & 20 & 60 & 60 & 30 & 290\\ 
QAM & 120 & 5 & 60 & 30 & 5 & 220\\ 
QAM & \textbf{15} & \underline {180} & 20 & \textbf{3} & 10 & 228\\ 
\midrule 
Median & 120 & 30 & 60 & 30 & 30 & 265\\ 
\bottomrule 
\end{tabular}

\end{table}

Based on responses of difficulty rates, in general, participants did not meet much difficulty in applying the tool and analyze its outputs.
Within these two experiments, none of the participants rated \emph{Very Difficult} in any of the five steps.
In the User Study 2021,
the participant met some dependency conflicts when using \evo client library.
This required a manual fix, and therefore the participant rated this task as \textit{Moderate}.
%
In the User Study 2023, 
with 10 participants on five tasks (in total 50 rate answers), there are two \emph{Difficult} rates on setup and three \emph{Difficult} rates on writing driver.
 For the rest of the answers, they are either neutral (i.e., \textit{Moderate}) or positive (i.e., \emph{Easy} or \emph{Very Easy}).
Regrading setting up \evo, two participants met the same difficulty when building it from source code as they needed to solve versions of tools and dependencies, such as Kotlin, Java 8/17 and Maven.
Regarding writing driver, its rate might depend on familiarity of participants on the implementation of the SUT.
For instance, three participants (all \emph{QA Engineer}s) considered that it is \emph{Difficult} to implement methods for starting/stopping/resetting the SUT programmatically, and
we received two \emph{Very Easy} answers from a \emph{Development Manager} and a \emph{Developer}.
If the participants (such as QA Engineer) do not need to write the code for programmatically handling the SUT (e.g., write unit tests or manual tests), they might meet some difficulties to figure out how to do that.
Regarding rates of resolving dependencies and running \evo, 70\% participants considered it \textit{Moderate} and 30\% participants considered it \textit{Very Easy}.
Executing tests seems be evaluated as the easiest task as 80\% participants rated it either \emph{Easy} or \emph{Very Easy}, where only 20\% participants rated it \emph{Moderate}.

Table~\ref{tab:appExTime} presents the time cost spent at each step per participant.
The time cost varies from participant to participant, and
there are various factors which can impact results, e.g., their daily task and proficiency relating to techniques that \evo employs or needs.
Those who spent less time (e.g., DPE and DPM took 39 minutes) might be familiar with tasks such as resolving dependencies with Maven (Resolve dependency), starting/stopping/resetting APIs programmatically (Write driver), starting the tool with the command line (Run Evomaster), and executing JUnit tests independently of their testing framework (Execute tests).
The 
median cost is 265 minutes
For instance, the minimum time cost of applying \evo and executing generated tests is 39 minutes, while the maximum time cost is 720 minutes (i.e., 12 hours $\approx$ 2 working days).
Based on median time cost for the tasks, \emph{resolving dependencies}, \emph{running the tool} and \emph{executing tests} are tasks which take less time (i.e., 30 minutes), while setting up \evo needs more time to figure it out (i.e., 2 hours).

Note that the time values reported in Table~\ref{tab:appExTime} are \emph{self-reported}.
Getting exact time measures would had significantly complicated the study, as there would had been the need to monitor each of the participant's laptop/PC.
And those machines are locked-down to protect the IP of Meituan (e.g., source-code) from cyber threats (i.e., we would not had been able to install any monitoring software on such machines, as that would had been a security vulnerability).
 As the exact time values are not critical, and approximate self-reported values could be enough to get insight into this problem, we considered such an approach as a reasonable compromise.

Regarding tool setup, in 
the two user studies, we found out that surprisingly all participants did set up \evo by building it from source code.
The choice of setup might relate to documentation, that might lack step-by-step guideline to get started with \evo.
In addition, the documentation of \evo is currently written only in English.
As all participants are not native English speakers, then they might not  take full advantage of all installation information when they get started to use the tool (e.g., video tutorials).

\textbf{Results outside of Meituan.}
Out of the five practitioners who participated in the User Study 2024, only one (i.e., Software Architect (SWA)) successfully completed all five steps and was able to use \evo for test generation, while the other four did not manage to write a driver within their free time.
Answers provided by the one who completed all steps are:
\begin{itemize}
	\item difficulty rates: \emph{Moderate} for setting up, resolving dependencies, running the tool and executing tests, and \emph{Difficult} for writing driver;
	\item time cost: 10 minutes for setting up, 5 minutes for resolving dependencies, 120 minutes for writing driver, 10 minutes for running the tool, and 1 minute for executing tests.
\end{itemize}

Results indicate that among the tasks required for utilizing \evo, writing drivers to enable white-box testing appears to be the most challenging.
Compared to results obtained at Meituan, writing drivers also received the most \emph{Difficult} rates (see Figure~\ref{subfig:appEx2023Rates}), and it stands out as the second most time-consuming task for participants (see Table~\ref{tab:appExTime}).

\textbf{Discussion.}
These results provide some interesting insight and reflections.
First, who are the target users for these fuzzers in industry?
Developers and testers are likely candidates, but their background profiles can be very different.
A software developer would have the coding skills to setup the drivers to enable white-box testing, whereas a tester might not, or require significantly more time.
For these latter, a black-box approach that requires no coding setup might be more effective, as easier to use.
Even if white-box testing can achieve significantly better results (i.e., higher code coverage and fault detection), there is still an important role left for black-box testing.
Notice that, at the time of this writing, for RPC APIs we only support white-box testing in \evo.

Another interesting aspect is regarding possible language barriers.
The de-facto language of science is English, and most software developers in the world have to deal with the English languages on a daily base (e.g., most programming languages are text based, and use keywords from the English language).
This can create a bubble, where academics could wrongly assume that the output of their work written in English would be understandable to practitioners in industry at large, and not possibly just to a minority that can read and understand English.
It is not in the scope of this paper to do a full analysis of language barriers in the software industry around the world, but it is clearly an issue that we experienced in this work, especially when dealing with ``non-coders''.

Research prototypes are already seldom documented.
Providing documentation and user manuals in different languages (e.g., Mandarin and Spanish) would not be viable in most cases.
However, that could be yet another barrier for technology transfer from academic research to industrial practice.

Based on results of usability obtained in the three user studies, writing drivers appears to be the most challenging task. 
This may be because it requires additional skills, such as familiarity with how to start/stop/reset SUT programmatically, and programming proficiency.
To improve the usability of \evo, the task should be simplified, and better documentation (e.g., better tutorials) should be provided.
It is also worth mentioning that, 
our industrial partner now have automated the five steps of applying it on their services and integrated \evo into their testing pipelines.
Industrial APIs are often built with the same pattern and use dependencies from their internal repository, thus, 
code for writing the driver (such as starting, stopping and resetting) and configuration of resolving dependencies can be templatized.
In their testing pipeline, \evo is pre-setup with a specific version, then it could be further used for fuzzing industrial APIs automatically.
Settings (such as choice of search budget and using specific algorithm options) for generating tests with \evo can be configured easily with a script, and our industrial partner uses the same setting for all its APIs.

In the User Study 2024, we received fewer valid responses (i.e., $1/5=20\%$) compared to valid responses obtained at Meituan (i.e., $11/11=100\%$).
This discrepancy might depend on the context, i.e., if the research collaboration exists. 
There, the user studies at Meituan were approved by higher management, and participants worked on it during their working hours.
In monetary value, that was a non-negligible cost (e.g., the salary of the employees).
This was not the case in these other five enterprises, where there is currently no research collaboration with the authors of \evo.
Collaborations between academia and industry are valuable, but hard to setup and maintain~\cite{garousi2016challenges,garousi2019characterizing}.

Learning to use a new tool can take time.
But, once got familiar with it, each following use would be easier and less time consuming.
However, for practitioners there is the need to see some direct benefits in it before being willing to invest time to learn how to use such a new tool, or build a customized infrastructure to easy its use on their systems (as done at Meituan).

\begin{result}
	\textbf{RQ1.1}: To apply \evo at Meituan,
	based on difficulty rates from 11 participants on five steps, it is likely not problematic to use \evo as we received 0 very difficult rate, 5 difficult rates (9\%), 23 moderate rates (42\%) and 27 easy/very easy rates (49\%), and the median time cost to apply the tool with the five steps for the first time is 265 minutes.
	In the user study outside of Meituan involving five participants from different companies, we observed that writing drivers appears to be the most challenging task, with four out of five participants struggling to use \evo and failing to complete the task.
	Similarly, this task was also rated as the most Difficult in user studies conducted at Meituan.
	To enhance usability, it requires simplification and improved documentation.
	As the industrial APIs are often built with same pattern, applying \evo has been automated, and now \evo has been integrated into the development pipelines at Meituan.
\end{result}

\subsubsection{Results for RQ1.2}
\label{subsec:rq12}

\evo can be applied for automating test case generation for industrial APIs.
 But, in order to better support such automation, there still exist some challenges which require both researchers and industrial partners to address.
With the first experiment using \evo \vfirst in 2021,
we found that, since \evo mainly addresses REST APIs and did not directly support any RPC mechanism (e.g., such as Apache Thrift) yet, in order to employ \evo on their services, our industrial partner needed to implement an additional REST layer.
 This was implemented based on  one-to-one mapping from each RPC function to a REST endpoint. Building
such a layer was rather straightforward.
However, the manual work with the additional layer might bring new problems, and without a native RPC support, the effectiveness of \evo might be limited (e.g., strategies specific to REST domain might not work for the services developed with RPC).
Therefore, we developed a native white-box fuzzing for RPC APIs~\cite{zhang2023rpc} (supported in \evo \vsecond~\cite{zenodo161evomaster}) that is currently deployed in the testing pipelines at Meituan.

Another challenge relates to reset the state of SUT in  industrial setting.
In order to generate independent tests, \evo requires a proper reset of the SUT, e.g., clean modified data in databases, reset/manipulate states of related services.
The industrial APIs often are parts of a large microservice architecture, and they require to interact with databases and other external services (see Figure~\ref{fig:sut}).
At Meituan, they had deployed a specific testing environment where all services are up and running (which is a typical practice in industry).
Within this environment, a service under test is able to interact with all its required services (e.g., external services and databases shown in Figure~\ref{fig:sut}).
With such a real industrial setting,
it is not trivial to reset the state of the SUT.
First, the interacted external services are not controllable from the test cases, which means that we cannot manipulate them to specific states before executing a new test.
%
Besides, the databases in real practice are more restricted and complex.
For instance, in the first experiment, we found that some types of SQL commands are blocked (e.g., \texttt{TRUNCATE}), then it is not applicable to reset the databases using SQL commands, such as a \texttt{DbCleaner} utility provided in \evo to clean data in directly connected SQL databases.
Once the SQL command restriction issue was identified, we discussed with our industrial partner, and they implemented a solution to provide us access to execute SQL commands.
But a challenge we faced later is how to handle a large amount of data in the databases.
In industry, to better test their services, they often prepare data in the database, and the amount of data can be very large from the point of view of testing (e.g., 1,017,504 rows of data in \csE shown in Table~\ref{tab:sut-info}), which could be created manually or collected during production.
Such data is maintained and shared to all development and testing groups.
To manipulate various states of the SUT,
\evo integrates SQL handling to insert data into the databases with defined SQL heuristics for test generations~\cite{arcuri2020sql}.
Considering the large amount of data in the databases, it is not viable to reset them (i.e., clean and then re-add) before executing every test, as it would become a drastic performance bottleneck.
However, if we do not clean the data, the newly inserted data might affect their testing environment.
In a scientific research context, our industrial partner could help us to isolate a testing environment with an empty database (the empty database can significantly reduce the time cost of the reset), as we did in~\cite{zhang2023rpc}.
But, it is not a feasible solution for our industrial partner when they apply \evo in their real development/testing process.
For instance, isolating a complete testing environment is time-consuming, and it is impractical to perform the operation every time when applying \evo.
In addition, with enabled SQL handling, the tests generated by \evo would have \texttt{INSERT} SQL commands and perform the reset that likely introduces problems into their testing environment.
As discussed with our industrial partner, without yet a viable solution to handle such large amount of data, we decided to disable the reset and SQL handling in the real application of \evo at Meituan.

In the second experiment we conducted in 2023, \evo was configured with the settings used by our industrial partner, i.e.,  without the reset and SQL handling.
To avoid conflicts with other teams which might test the same services, our industrial partner isolated a testing environment where all connected services are up and running, and databases have the same data as the main testing environment for running \evo.
In this isolated environment, the databases and services (the SUT and its connected services) cannot be accessed by other teams.
Thus, the tests generated by \evo could be further used in the main testing environment.
Note that there might be side-effects when disabling the database reset, e.g., a sequence of executing tests might affect the results of other tests.
However, such possible side-effect was deemed acceptable by our industrial partner.



Furthermore, we found that, besides authentication restrictions, there also exist some restrictions based on intra-service business logic.
For instance, an endpoint in a SUT \emph{X} might only accept requests which contain specific information, and the specific information could be put by an endpoint in another service \emph{Y}.
Thus, in order to access the endpoint in the SUT \emph{X}, the endpoint in \emph{Y} must be invoked  first, or the request must be sent by the endpoint in \emph{Y}.
Such info among services are handled in an internal \textit{Tracer} service developed  by Meituan.
Thus, in order to access the endpoints in testing, we extended \evo for providing a method for users to define 
pre-actions to resolve such constraints,
 and the invocation of the pre-actions would be handled with a probability as authentication in \evo, i.e., a probability to control whether the request contains such required information.
The authentication is currently configured automatically in the driver generated by Meituan, using a default setting.

 	
\begin{result}
	\textbf{RQ1.2}:
	The challenge of having a native support for fuzzing RPC APIs has been addressed with the latest version of \evo.
	However, due to complex interactions with databases and external services, there exist challenges in having an effective solution to properly reset the state of the SUT and enable SQL handling.
\end{result}

\subsection{Results for RQ2 : Effectiveness}
\label{subsec:rq2}

Based on results obtained by \evo on industrial APIs, the effectiveness of \evo was firstly analyzed by the authors of \evo (Section~\ref{subsecsec:rq2s1}).
The analyses were performed to investigate the performance of \evo on the industrial APIs with various time budget settings, and to assess the line coverage and fault detection achieved by \evo on the given industrial APIs.
However, for the User Study of 2024, due to the lack of collaborative agreements with the enterprise, we do not have access to the code and results, and therefore, cannot perform the analyses.

The effectiveness of \evo was also assessed by practitioners with tests generated by \evo in terms of readability and quality (Section~\ref{subsecsec:rq2s2}).
In the user studies at Meituan, we provided the tests which achieved the best results for each SUT and results of the analyses (such as line coverage and number of potential faults obtained by \evo with each time budget) to participants for conducting the assessment.
For the user study outside of Meituan, participants can select the tests by themselves for the assessment.

\subsubsection{Results for RQ2.1}
\label{subsecsec:rq2s1}

\begin{figure}
	\begin{subfigure}{.33\textwidth}
		\includegraphics[width=\linewidth]{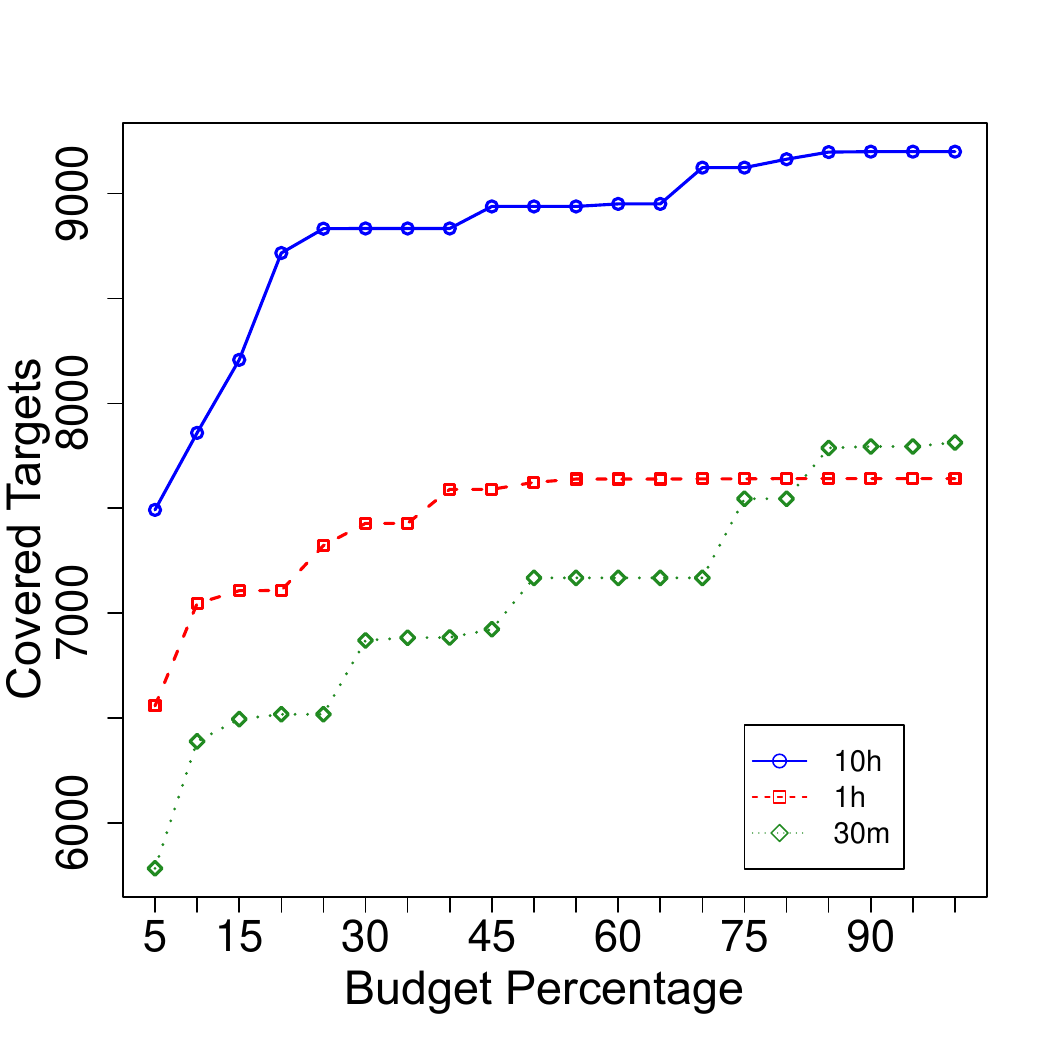}
		\caption{\csA}
		\label{subfig:cs1}
	\end{subfigure}\hfill
	\begin{subfigure}{.33\textwidth}
		\includegraphics[width=\linewidth]{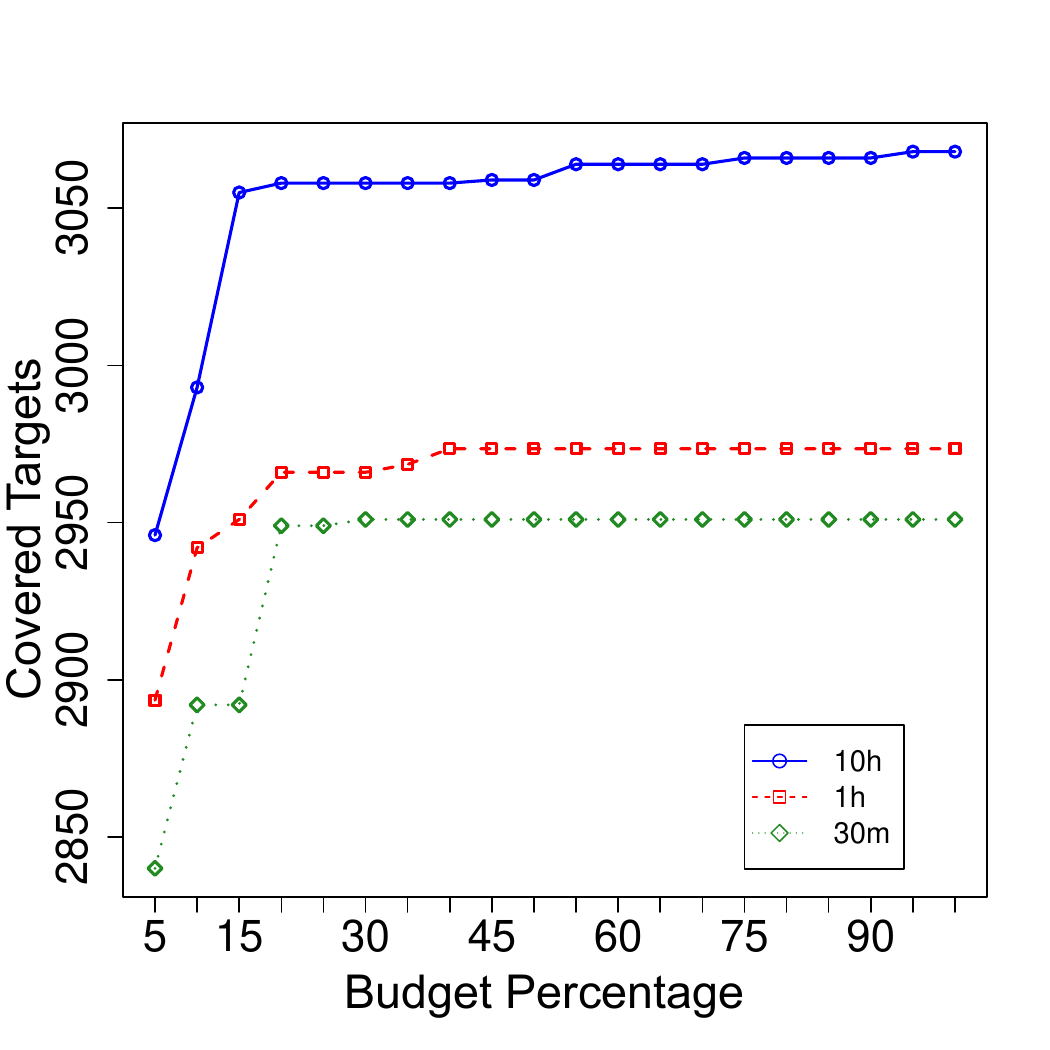}
		\caption{\csB}
		\label{subfig:cs2}
	\end{subfigure}\hfill\\
	\begin{subfigure}{.33\textwidth}
		\includegraphics[width=\linewidth]{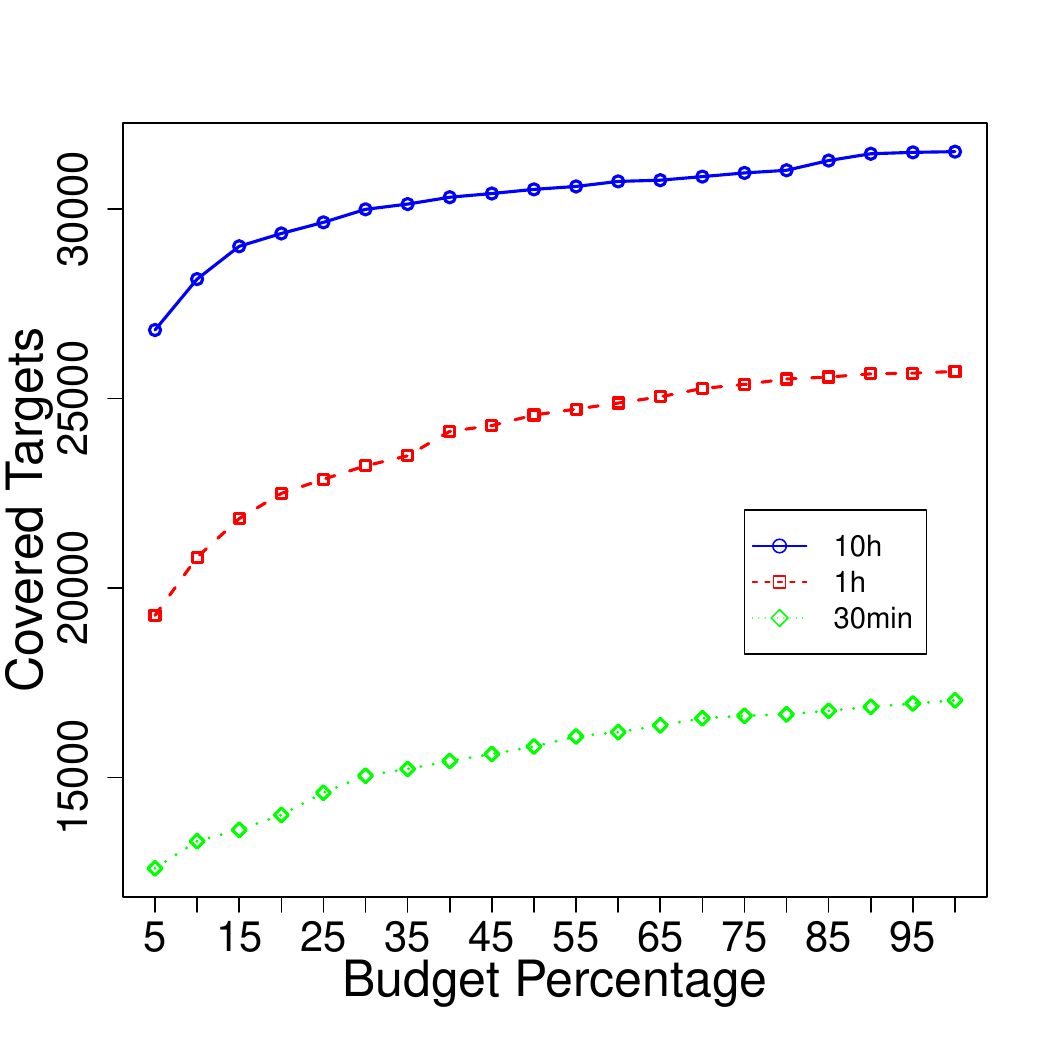}
		\caption{\csC}
		\label{subfig:cs3}
	\end{subfigure}\hfill
	\begin{subfigure}{.33\textwidth}
		\includegraphics[width=\linewidth]{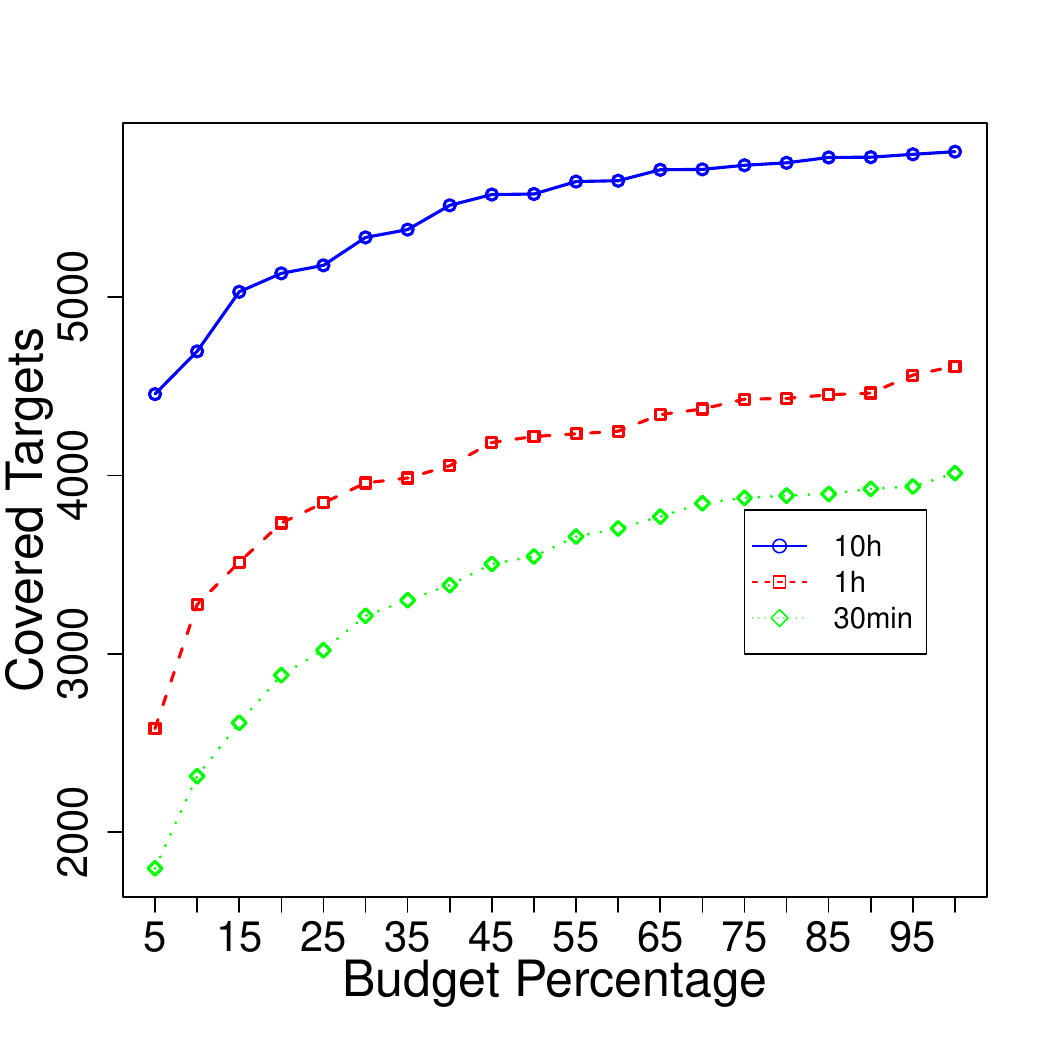}
		\caption{\csD}
		\label{subfig:cs4}
	\end{subfigure}\hfill
	\begin{subfigure}{.33\textwidth}
		\includegraphics[width=\linewidth]{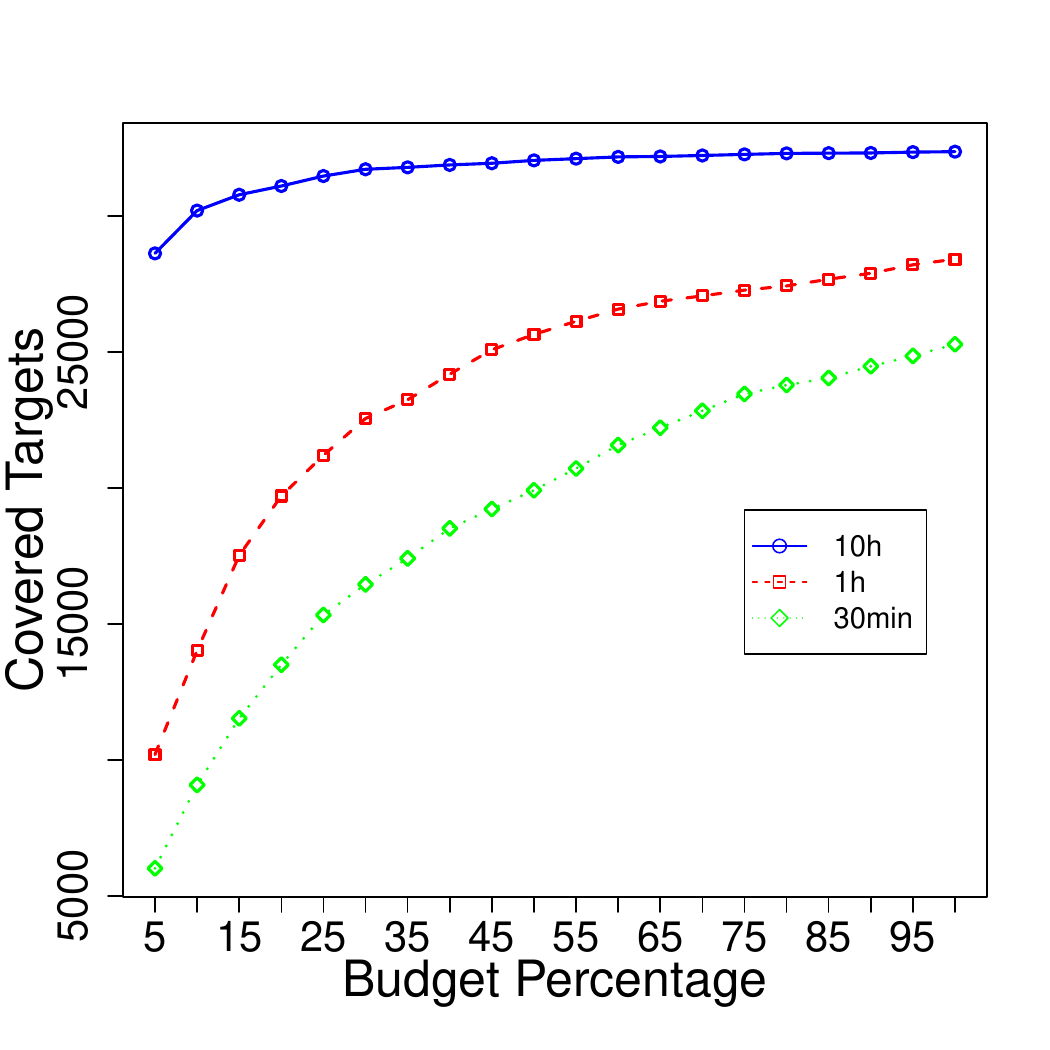}
		\caption{\csE}
		\label{subfig:cs5}
	\end{subfigure}\hfill
	\caption{Average covered targets (y-axis) with 30m ({\color{ForestGreen}green} line), 1h (\textcolor{red}{red} line) and 10h (\textcolor{blue}{blue}  line) throughout the search for industrial applications, reported at 5\% intervals of the used budget allocated for the search (x-axis)}
	\label{fig:rq2-targets}
\end{figure}

\textbf{Performance of \evo on industrial APIs with various time budget settings}.
\evo identifies various metrics related to testing criteria as testing targets, such as code coverage (e.g., class, method, branch, statement), identified faults and domain specific targets (e.g., different execution results for RPC), and produces tests in order to maximize the number of targets being covered. 
For instance, a line of code such as  \texttt{if(x == 42)} will be treated as three distinct testing targets for the line itself (denoted as LINE), the branch where the condition is evaluated as true (denoted as BRANCH\_TRUE), and the branch where the condition is evaluated as false (denoted as BRANCH\_FALSE) respectively.
Once this line is executed, we consider the LINE target is covered, and the two branches targets are \emph{reached}.
The BRANCH\_TRUE target can be considered as \emph{covered} only if \texttt{x} is 42, while the BRANCH\_FALSE target can be considered as \emph{covered} only if \texttt{x} is not 42.
In order to cover these three targets, tests need to be generated to execute the line and to cover true and false outcomes of the branch.
More detail about the testing targets can be found in~\cite{arcuri2018test,zhang2023rpc}.
To investigate performance of \evo with various time budget setting,
Figure~\ref{fig:rq2-targets} plots the number of testing targets covered at every 5\% budget used by the search for the five applied SUTs, with search budgets of 30 minutes, 1 hour and 10 hours.

Based on the figures,
for the SUTs, results achieved by 10 hours clearly outperform results with 30 minutes and 1 hour.
Comparing results between 30~minutes and 1 hour, their performances are close except \csA (see Figure~\ref{subfig:cs3}).
In \csA, the budget with 30~minutes achieves even better results than the 1-hour budget.
Considering the random nature of search algorithms, it could happen by chance as the experiment on \csA was only performed once. But
it also reveals that probably there is a lack of exploration of the search landscape with the small budget.

Applying search for many objectives such as test generation, it is essential to explore the search landscape at early stages for providing diversification of individuals.
Then, such individuals could enable a further possibility to cover new targets, e.g., code coverage and fault finding in our context, in the later exploitation stage.
Therefore, with automated testing approach such as \evo, a longer budget would be required, i.e., more than 1 hour, to achieve better results.
However, a too long search budget (e.g., 24 hours) might be not viable, if engineers are not willing to wait so long to get results.
But this also strongly depends on whether tools like \evo would be typically run on a developer machine, or on a remote dedicated CI server.
Also, it depends on whether the fuzzing is done during a regular development day, or before a major software release (this latter would like entail spending a longer time for testing).

\textbf{Line coverage achieved by \evo on industrial APIs.}
Table~\ref{tab:rq2-coverage} represents line coverage 
achieved by two versions of \evo (i.e., \vfirst and \vsecond) in the two experiments.
Results show that \evo \vfirst achieved on average 28.9\% line coverage in \csA and \csB, and \evo \vsecond achieved on average 18.1\% line coverage in \csC, \csD and \csE.
It would be interesting to study performance between two versions of \evo if possible.
But, as seen descriptive statistics of the SUTs in Table~\ref{tab:sut-info}, industrial SUTs could be significantly updated within 1.5 years (i.e., \csC has more 24,816 LOC than \csA, and \csD has more than 14,201 LOC than \csB), and these SUTs can be considered as different ones.
Therefore, we cannot compare the two versions of \evo based on the results achieved on \emph{CS1} and \emph{CS2}.
In  industrial settings, just for research purposes cost-wise it is not viable to run the  old versions of the services as their connected services and databases would need to be downgraded as well.
However, as this study is aimed at investigating \evo in a real industrial context from the viewpoint of practitioners, comparing two versions of \evo are out scope of this paper.

In User Study 2021,
to promote further integration of our approach into industrial setting,
we performed a manual review on the tests generated with 10 hours and coverage reports with the industrial partner, together.
Such review could help us to understand about industrial APIs and provide industrial partner info about potential benefits (if any) achieved by our approach.
By reviewing coverage report, classes for implementation industrial services at Meituan 
can be categorized into three parts, i.e., \emph{Client} for designing how the RPC-based API can be accessed, \emph{DAO} for data persistence and communication with database, and \emph{Server} for implementing business logic.
Line coverage achieved by \evo on each part is (1) \csA: 62.8\% on \emph{Client}, 10.9\% on \emph{DAO}, and 46.3\% on \emph{Server}; (2) \csB: 79.8\% on \emph{Client}, 15.3\% on \emph{DAO}, and \emph{57.1\%} on \emph{Server}.
First, as seen from the line coverage on each part, we found that achieved coverage on \textit{DAO} is low.
One possible reason for the low coverage in \textit{DAO} is that we disabled \textit{SQL heuristics} in these experiments, as discussed in Section~\ref{subsec:rq12}.
Thus, without any heuristic on SQL, it might result in such underperformance in \textit{DAO}.
In addition, we also found that there exist some unreached classes in \textit{DAO} which define SQL operations (e.g., \texttt{SELECT ALL}, \texttt{DELETE BY WHERE}), and those classes seem never been used (i.e., dead-code).
As checked with the developers at Meituan, those classes are actually automatically generated by their development framework.
In addition, they also pointed out that, in their services, there possibly exist some other unused code which is not deleted due to their frequent updated services.
We also compared to manually written automated end-to-end (E2E) tests, if the SUT has any, in terms of code coverage.
At the time of performing the first experiment in 2021, there were 11 manually written JUnit tests for \csB which achieve 16.8\% line coverage, and no manually written JUnit tests for \csA.
Our generated tests (i.e., 24.8\% with 30m, 26.9\% with 1 hour, and 32.6\% with 10 hours in Table~\ref{subtab:coverage2021}) show clear better results than the existing manually written JUnit tests.

By reviewing the tests,
the industrial partner found that, 
the tests outputted by \evo are quite useful for them from three perspectives which are rarely covered with manual testing and manually written automated tests\footnote{
	At Meituan, there exist two kinds of system-level tests.
	One is manual testing, i.e., testers act as their real users from the user side involving many services for processing the user request (see Figure~\ref{fig:sut}) as real business scenarios.
	The other kind is manually written automated tests with JUnit, which
	mainly addresses their business scenarios and logic for a service.
}:
(1) \textit{Input validation}: the generated tests are capable of covering various input validations of the services, and their combinations;
(2) \textit{Assertion generation for complex responses}: \evo is able to generate successful requests to the SUT, and have complete assertions with respects to the responses that could be not possibly achieved  (way too time consuming) by manually written tests when the elements in the response are many (e.g., large and complex returned responses);
(3) \textit{Exceptional branch coverage}: the tests also enable covering different critical exceptional scenarios such as invalid operations on resources at a given state, authentication, or time related constraints on operations (e.g., exceptions regarding too frequent operations on the endpoints).
It is noteworthy that the identified
benefits gave our industrial partner confidence to take effort to use our approach and promote further its integration into their development/testing process.

In \evo \vsecond, we supported a native fuzzing of RPC APIs~\cite{zhang2023rpc}.
As the issues relating database handling are not tackled yet, effectiveness  of \evo in terms of line coverage is still limited as results obtained in User Study 2023.
By looking at
code coverage in the two user studies (see Table~\ref{tab:rq2-coverage}), \evo might achieve better results on APIs which have relatively less complexity.
For instance, for \csA (32,292 LOC, and 33 endpoints), \csB (12,152 LOC, and 13 endpoints) and \csD (26,353 LOC, and 29 endpoints), \evo achieved around 30\% line coverage using 10-hours time budget.
For \csE (193,024 LOC and 63 endpoints), \evo achieved 22.8\% on average of 10 runs using 10-hours time budget.
On \csC (57,209 LOC and 62 endpoints), the performance of \evo is limited, i.e., 10.4\% on average.
Compared to \csE, \csC has less code and similar amount of exposed endpoints, but, \evo achieved the worst code coverage.
Then, by further investigating \csD and discussing it  with the industrial partner, we found that this service involves much more complex business logic than \csE, which might explain the underperformance of \evo.
Note that as parts of user studies, analyses in terms of effectiveness (such as line coverage) aimed to provide evidence of potential benefits \evo can bring to industry then promote further integration of our approach into industrial setting. 
Analyzing the limitations of \evo in fuzzing industrial APIs is beyond the scope of these user studies that requires the effort of a dedicated paper, e.g.,~\cite{zhang2023open}. 


\begin{table}
	\centering
	\caption{Results of line coverage and detected faults achieved by \evo with three time budget settings, i.e., 30 minuses, 1 hour and 10 hours.}
	\label{tab:rq2-coverage}
	
	\begin{subtable}[h]{1\textwidth}
		\centering
		\small
		\caption{Results of one run collected in 2021 using an additional REST API layer. Note that \# Detected Faults (REST) was reported by \evo based on 500 status code or mismatched schema for REST API, and \# Detected Faults (Manual) was based on a manual review performed by the first author and one employee at Meituan.
		}
		\label{subtab:coverage2021}
		\begin{tabular}{ll | rrrr}
			\toprule
			SUT &  TB & \#Tests Reviewed & Line Coverage \% & \# Detected Faults & \# Detected Faults \\
			&   & Manually & & (REST) & (Manual) \\
			\midrule
			\csA& 10h &231 &33.5  & 12 & 34\\
			& 1h &  &27.1  & 6 & \\
			& 30m &  &27.8  & 3 & \\
			\csB&  10h& 64&32.6 & 13 & 18\\
			&  1h& &26.9 &  8 & \\
			&  30m& &24.8 & 8 & \\
			\midrule 
			\multicolumn{2}{l}{Arithmetic Mean}  & &28.9 & 8.3 & 26 \\ 
			\bottomrule
		\end{tabular}
	\end{subtable}
	
	\bigskip
	
	\begin{subtable}[h]{1\textwidth}
		\centering
		\small
		\caption{
			Results of 10 runs collected in 2023. The results are reported with mean, median, minimum and maximum of 10-times repetition for each setting on each SUT as \texttt{Mean} (\texttt{Median}) [\texttt{Minimum}, \texttt{Maximum}].
		}
		\label{subtab:coverage2023}
		\begin{tabular}{ ll |  rr}\\ 
\toprule 
SUT & TB  &Line Coverage \% & \# Detected Faults  \\ 
\midrule 
\emph{CS1-2023}&10h & 10.4 (10.3) [9.9, 11.6] & 188.0 (188.5) [170, 210] \\ 
&1h & 8.7 (8.6) [8.0, 9.6] & 73.2 (73) [63, 82] \\ 
&30m & 5.8 (7.8) [0.2, 8.8] & 46.8 (64) [1, 76] \\ 
\emph{CS2-2023}&10h & 30.6 (30.8) [27.9, 33.4] & 23.4 (23.5) [22, 25] \\ 
&1h & 24.5 (24.7) [22.5, 26.8] & 21.3 (22) [20, 23] \\ 
&30m & 21.6 (21.4) [17.7, 26.2] & 18.9 (20) [12, 22] \\ 
\emph{CS3-2023}&10h & 22.8 (22.8) [22.4, 23.4] & 72.7 (72.5) [71, 78] \\ 
&1h & 20.2 (20.1) [19.7, 21.1] & 71.6 (71) [69, 80] \\ 
&30m & 18.0 (18.1) [17.0, 19.0] & 67.7 (68) [66, 69] \\ 
\midrule 
\multicolumn{2}{l}{Arithmetic Mean} & 18.1 & 64.8 \\ 
\bottomrule 
\end{tabular} 

	\end{subtable}
	
\end{table}

\textbf{Faults detected by \evo.}
	Results of fault detection in the two user studies at Meituan are represented in Table~\ref{tab:rq2-coverage}.
For \csA and \csB, 
the results reported by \evo \vfirst (\# Detected Faults (REST)) are 
derived based on REST characteristics, i.e., 500 status code and mismatch issues in OpenAPI schemas.
To avoid potential issues of identifying faults in RPC APIs using domain knowledge of REST,
we performed a further manual fault identification with an employer who is familiar with two APIs at Meituan using the generated tests (i.e., execute the tests and review the tests and their execution results).
	By reviewing 231 tests for \csA and  64 tests for \csB,	
	34 unique faults for \csA and 18 unique faults for \csB with the manual identification are reported in the \# Detected Faults (Manual) in Table~\ref{subtab:coverage2021}.
	Note that each test generated by \evo is composed of not only requests to the SUT but also the corresponding responses. 
	Thus, the employee can identify faults by checking whether the responses match the expected outcomes based on the requests or whether exceptions are properly handled.  
With the first experiment, we found that, without a native support on RPC-based API fuzzing, it is not effective to recognize faults in tests generated by \evo.
Such problem has been solved 
in the second experiment with \evo \vsecond as we have enabled such a native support for automated fuzzing of RPC API that comprises a schema parser for RPC API and new heuristics specific to RPC domain (such as covering various responses of RPC calls and maximizing found faults in RPC APIs)~\cite{zhang2023rpc}.
	For \csC, \csD, and \csE, with \evo \vsecond, on average 64.8 faults (see \# Detected Faults in Table~\ref{subtab:coverage2023}) are derived based on RPC characteristics, i.e., exceptions thrown from distinct places in the code or service error derived using customized methods~\cite{zhang2023rpc}.
Ideally, all such problems should be fixed, e.g., exceptions should be properly handled to avoid further propagation in the microservices architecture.
	Hence, there is no need to conduct the manual fault identification for results reported by \evo \vsecond.

\begin{figure}
\begin{lstlisting}[language=java]
public Response interfaceName(Reuqest resuest){
	Preconditions.checkArgument(resquest.getuserId()>0,"invalid userId");
	...
	return response
}
\end{lstlisting}\hfill
\begin{lstlisting}[frame=single,numbers=none]
exceptionName:com.xx.aa.bb.cc.thrift.common.XXXXXTException
exceptionMessage:invalid userId
\end{lstlisting}
\caption{An example of a low-priority fault to fix}
\label{fig:fault-low}
\end{figure}

\begin{figure}
\begin{subfigure}{1\textwidth}
\begin{lstlisting}[language=java]
public Response interfaceName(Request request){
	List<ResponseDo> list  = service.queryData(request.getIds());
	List<Integer> ids = list.stream().map(ResponseDo::getId).collect(Collectors.toList())
	ResultSet rs  = rep.findDetailsByIds(ids);
	...
	return response;
}
\end{lstlisting}\hfill
\begin{lstlisting}[frame=single,numbers=none]
exceptionName:java.sql.SQLException
exceptionMessage:Error querying database. 
Cause: java.sql.SQLException: java.sql.SQLException: java.lang.ArrayIndexOutOfBoundsException\n
### The error may exist in class path resource [mybatis/shard/XXXXMapper.xml]\n
### The error may involve defaultParameterMap\n
### The error occurred while setting parameters\n
### SQL: select xxxxx from XXXX where xxx in ( ? , ? , ? , ? )\n
### Cause: java.sql.SQLException: java.sql.SQLException: java.lang.ArrayIndexOutOfBoundsException\n; 
uncategorized SQLException; SQL state [null]; error code [0]; java.sql.SQLException: java.lang.ArrayIndexOutOfBoundsException; nested exception is java.sql.SQLException: java.sql.SQLException: java.lang.ArrayIndexOutOfBoundsException
\end{lstlisting}

\caption{A high-priory fault before fixing}
\label{subfig:before-fix}
\end{subfigure}\hfill
\begin{subfigure}{1\textwidth}
\begin{lstlisting}[language=java]
public Response interfaceName(Request request){
	Response response = null;
	List<ResponseDo> list  = service.queryData(request.getIds());
	List<Integer> ids = list.stream().map(ResponseDo::getId).collect(Collectors.toList())
	// fix the fault by adding an empty check for ids
	if(ids.isEmpty()){
		return response;
	}
	ResultSet rs  = rep.findDetailsByIds(ids);
	...
	return response;
}
\end{lstlisting}
\caption{A high-priory fault after fixing}
\label{subfig:after-fix}
\end{subfigure}
\caption{An example of a high-priority fault to fix}
\label{fig:fault-high}
\end{figure}


	Regarding fault fixing, in the first experiment, out of the total  detected 52 faults (i.e., $34+18$), the industrial partner treated 21 faults (10 for \csA and 11 for \csB) as relatively critical. 
	These 21 faults have been fixed.
	In the second experiment, with a setting of 10-hours time budget, a total of 2,841 faults were detected over 10 repetitions (i.e., $188.0 \times 10$ for \csC, $23.4 \times 10$ for \csD, and $72.7 \times 10$ for \csE).
	It was impractical to check all of them.
	However, since \evo now has been integrated into the pipeline at Meituan, the fault fixing rate is around 10\% reported by our industrial partner.
	The fixing rate might relate to the priority of the faults to be fixed and the effort required to check them.
	For instance, Figure~\ref{fig:fault-low} represents a fault detected by \evo \vsecond which our industrial partner considered low-priority to fix, while Figure~\ref{fig:fault-high} represents a high-priority fault to fix.
	In Figure~\ref{fig:fault-low}, a \texttt{com.xx.aa.bb.cc.thrift.common.XXXXXTException} exception might be thrown in line~2 during request validation if the specified \texttt{userId} does not exist in the database.
	As the API is typically invoked by other APIs from the large microservices rather than being directly accessed by users, our industrial partner considers that it is uncommon to have invalid \texttt{userId}, then the fault is the low-priority to fix.
	In Figure~\ref{subfig:before-fix}, a \texttt{java.sql.SQLException} exception might be thrown in line 9 if the provided \texttt{ids} is empty, and how the fault is fixed is shown in Figure~\ref{subfig:after-fix}.
	Our industrial partner considers that exceptions related to the database are of high priority to fix, as the database may be accessed by many APIs, and \texttt{ids} could potentially be empty in the scenario.
	Moreover, the low fixing rate can also be attributed to the non-trivial effort required to check faults after each run of applying \evo.
	Since the faults are unclassified in \evo, QA engineers or testers must assess whether each fault requires attention before reporting them to developers.
	A further strategy is needed to better distinguish high-priority faults from all other faults.

\begin{result}
	\textbf{RQ2.1}: As a search-based approach, i.e., \evo, a relatively higher budget (e.g., 10-hours) is required to tackle industrial test generation.
	With the 
	recent \evo (i.e., \vsecond) on three industrial APIs at Meituan, it is capable of achieving up to 33.4\% (on average 18.1\%) code coverage and identifying up to 210 (on average 64.8) potential faults.
	The fault fixing rate is low (i.e., around 10\% reported by Meituan) that might relate to the the ratio of high-priority faults detected and non-trivial effort required to check all faults.  
\end{result}

\subsubsection{Results for RQ 2.2}
\label{subsecsec:rq2s2}

To evaluate effectiveness of \evo from the point of view of industrial practitioners, we conducted a questionnaire that comprises a set of questions (see Section~\ref{subsubsec:eff}) for collecting feedback of industrial participants about generated tests.
This questionnaire experiment was performed in all of the three user studies.
%

\begin{table}
	\centering
	\small
	\caption{Answers of pre-check questions of assessing effectiveness collected in User Study 2023 and User Study 2024}
	\label{tab:preffInfo}
	
	\begin{tabular}{ ll rrrrrr }\\ 
\toprule 
CS & Position & PR1: & PR2: Achieved  & PR2: Detected  & PR5: Time  & PR6: Enough \\ 
 &  & Familiarity & Code Coverage & Faults   &  Cost (minutes) &Time?  \\ 
\midrule
\multicolumn{4}{l}{\textbf{{User Study 2023 at Meituan}}}  & & &\\ 
\hline
 & QAM & 2 &  & \checkmark & 30 & \textbf{Y}\\ 
 & QAE & 5 &  & \checkmark & 60 & \textbf{Y}\\ 
\emph{CS1 2023} & QAM & 4 & \checkmark &  & 20 & \textbf{Y}\\ 
 & DPE & 5 &  & \checkmark & 5 & \textbf{Y}\\ 
 & QAE & 5 & \checkmark & \checkmark & 60 & \textbf{Y}\\ 
\midrule 
 & DPM & 5 & \checkmark & \checkmark & 120 & \textbf{Y}\\ 
 & PDM & 3 &  & \checkmark & 10 & \textbf{Y}\\ 
\emph{CS2 2023} & QAE & 5 & \checkmark &  & 20 & \textbf{Y}\\ 
 & QAE & 5 & \checkmark & \checkmark & 10 & \textbf{Y}\\ 
 & QAM & 5 & \checkmark &  & 10 & \textbf{Y}\\ 
\midrule 
 & DPE & 3 & \checkmark &  & 120 & \textbf{Y}\\ 
 & QAE & 5 & \checkmark &  & 120 & \textbf{Y}\\ 
\emph{CS3 2023} & QAE & 3 & \checkmark &  & 60 & \textbf{Y}\\ 
 & QAE & 5 & \checkmark &  & 120 & N\\ 
 & QAE & 1 & \checkmark &  & 120 & N\\ 
\midrule 
Median&  & 5& & &  60&  \\ 
Sum&  & &11 &7  &   & 13/15  \\ 
\hline \hline
\multicolumn{4}{l}{\textbf{{User Study 2024 outside of Meituan}}}  & & &\\ 
\hline
{Untitled} & {DPM} & {1} &  {\checkmark} &  & {-} & {N}\\ 
{Untitled} & {DPE} & {2} & {\checkmark}  & & {-} & {N}\\ 
{Untitled} & {SWA} & {4} &  {\checkmark} & & {30} & {N}\\ 
{Untitled} & {Tester/QAE} & {1} & &  {\checkmark}  & {-} & {N}\\ 
{Untitled} & {Tester/QAE} & {3} & &  {\checkmark}  & {-} & {N}\\ 
\midrule
{Median}&  & {2} & & &  {} &  \\ 
{Sum}&  & &{3} &{2}  &   & {0/5}  \\ 
\bottomrule 
\end{tabular}

\end{table}

Table~\ref{tab:preffInfo} represents responses for \emph{RP}s
(see Table~\ref{tab:allques}).
\emph{RP}s was designed after the first user study (i.e., User Study 2021) then employed in User Study 2023 and User Study 2024 in order to better understand answers of industrial participants, such as their familiarity with the API they employed in this user study (\emph{PR1}), the feature they preferred most in tests (\emph{PR2}), the property they consider most important in assessing tests in terms of readability (\emph{PR3}) and quality (\emph{PR4}), the time cost they spent to inspect the generated tests (\emph{PR5}), and whether they had enough time to conduct this user study (\emph{PR6}).

\prA
In User Study 2023 at Meituan, except 
one participant in \csC and one participant in \csE, all other participants have at least \emph{3-Moderate} familiarity.
Among the 15 participants, median familiarity is \emph{5-Very familiar}. 
In User Study 2024 outside of Meituan, among 5 participants from five different companies, two rated \emph{Moderate Familiar} and \emph{Familiar} while the other three rated either \emph{Very Unfamiliar} or \emph{Unfamiliar}.

\prB
In User Study 2023, among
the 15 participants, 7 out of them voted \emph{detected faults}, 11 out of them voted \emph{achieved code coverage}, none of them voted \emph{readability} and none of them specified other features they would most like to have.
Note that this question was marked to select only one choice.
However, there are three participants who selected two most important features (i.e., \emph{achieved code coverage} and \emph{detected faults}).
Based on the results, code coverage and fault detection seem the most two important features that participants would like to have in the  generated tests.
Based on the answers of these 15 participants, we received more votes for code coverage (i.e., 78.6\%) compared to fault detection (i.e., 46.7\%).
Responses obtained in User Study 2024 also have more votes for code coverage than fault detection, i.e., 3 \emph{vs.} 2.

\prC
In User Study 2023, we
received 7 answers relating to group tests, 5 answers relating to name of tests and 2 answers relating to provide comments.
All answers suggest to group the tests, name the tests or add comments that could refer to which RPC interfaces (see Section~\ref{subsec:subject}) involve.
If the test involves multiple interfaces, it would be better to represent what scenario the test examines.
In addition, two answers mention that tests should have clear requests, responses and test assertions.
Moreover, one suggests that the assertions would better to represent what it tests for, thus once it fails, it could help them to fix bugs.
Furthermore, there is one recommendation to better separate test data and test cases into different files, which might help them to analyze and maintain the tests.
Regarding the size of the tests, we received only one answer which concerns about this issue.
In User Study 2024, we received only one response, and the participant (i.e., DPM) considers grouping tests to be important for rating readability.

\prD
In User Study 2023,
13 participants out of 15 describe one of the most important properties to rate the quality of tests is related to real business scenario, e.g., whether the tests are able to test real business logic, whether the tests are able to test the core business logic, and what scenarios the tests are able to test (e.g., invalid input and happy path).
7 participants consider that high code coverage is important for them to rate the quality, and one participant considers the quality of assertions.
There is also one participant (who is a developer) considering the quality relating to the capability of tests to help developers to detect and fix bugs.
In User Study 2024, one (i.e., DPM) out of five participants answered this question and indicated that business logic coverage is important for rating the quality of tests.

\prE
Results obtained by employees at Meituan are reported in Table~\ref{tab:preffInfo}, i.e., $(0, 10m)$: 1, $[10m, 30m)$: 6, $1h$: 3, and $2h$: 5.
Median time spent to read and understand the tests is 1 hour.
However, as only SWA generated tests successfully with \evo in User Study 2024 (discussed in Section~\ref{subsubsec:rq11}), we only received one response, and the participant took 30 minuets to read and understand the tests generated by \evo.

\prF
In User Study 2023, among
the 15 participants, two of them claimed that they did not have enough time to read and understand the tests.
This could happen in the company as they have tight scheduled tasks for their daily jobs, and possibly they did not get enough time for this research task from their direct manager, or suddenly had other pressing work with higher priority.
Based on such responses, we excluded answers of these two participants on \csE.
Thus, for the experiment in 2023, the analysis of effectiveness was performed based on answers from 13 employees.
Regarding the responses in User Study 2024, all five participants claimed that they did not have enough time to inspect the generated tests. 
Given these results, it is not feasible for us to analyze assessments of the five participants outside of Meituan on tests generated by \evo (i.e., \emph{QB}s).
Conducting such as a time-intensive user study on effectiveness of a tool (such as, run the tool for fuzzing enterprise APIs with various settings of time budget, and take time to inspect results) is challenging, especially in companies with which the researchers do not collaborate.

\begin{table}
	\centering
	\small
	\caption{Answers of \emph{QB}1,2,5,7 and 8 in detail for each participant}
	\label{tab:effInfo}
	
	\begin{tabular}{ ll rrrrrr }\\ 
\toprule 
CS & Position & QB1: & QB2:  & QB5: Keep  & QB7: Time & QB8: Time\\ 
 &  & Readability & Quality  &  Tests? & Preference  & Selection \\ 
\midrule 
\emph{CS1 2021} & QAM & Low & Moderate & \checkmark &   &  \\ 
\emph{CS1 2021} & QAE & Low & Moderate & \checkmark &   &  \\ 
\emph{CS1 2021} & QAE & Low & Moderate & \checkmark &   &  \\ 
\midrule 
\emph{CS2 2021} & QAM & High & Moderate & \checkmark &   &  \\ 
\emph{CS2 2021} & QAE & High & Very High & \checkmark &   &  \\ 
\midrule 
\hline
\emph{CS1 2023} & QAM & High & Low & \checkmark & 2h & 10h\\ 
\emph{CS1 2023} & QAE & High & Low & \checkmark & 1h & 10h\\ 
\emph{CS1 2023} & QAM & Very High & Moderate & \checkmark & 2h & 10h\\ 
\emph{CS1 2023} & DPE & Very High & Very Low &  {$\times$}& default & 10h\\ 
\emph{CS1 2023} & QAE & Moderate & Moderate & \checkmark & 24h & 10h\\ 
\midrule 
\emph{CS2 2023} & DPM & High & Low & \checkmark & 1h & 10h\\ 
\emph{CS2 2023} & PDM & Low & Moderate & {$\times$} & default & 30m\\ 
\emph{CS2 2023} & QAE & High & Moderate & \checkmark & 1h & 10h\\ 
\emph{CS2 2023} & QAE & High & Moderate & \checkmark & 1h & 10h\\ 
\emph{CS2 2023} & QAM & Very High & Very High & \checkmark & 5h & 10h\\ 
\midrule 
\emph{CS3 2023} & DPE & Moderate & Low & \checkmark & 5h & 10h\\ 
\emph{CS3 2023} & QAE & Moderate & Moderate & \checkmark & 5h & 10h\\ 
\emph{CS3 2023} & QAE & Moderate & Moderate & \checkmark & 5h & 1h\\ 
\bottomrule 
\end{tabular}

\end{table}

\begin{figure}
	\centering
	\begin{subfigure}{1\textwidth}
		\includegraphics[width=.99\linewidth]{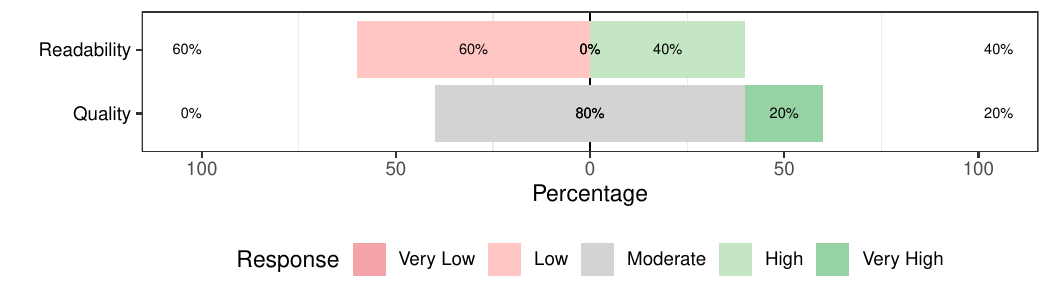}
		\caption{Diverging stacked barplot for responses of rates on readability and quality of generated tests using 5-points likert-scale collected in 2021. Note that positive responses (i.e., ``high'') extending to the right and negative responses (i.e., ``low'')  to the left from the diverging point.}
		\label{subfig:effEx2021Rates}
	\end{subfigure}\hfill
	\begin{subfigure}{1\textwidth}
		\includegraphics[width=\linewidth]{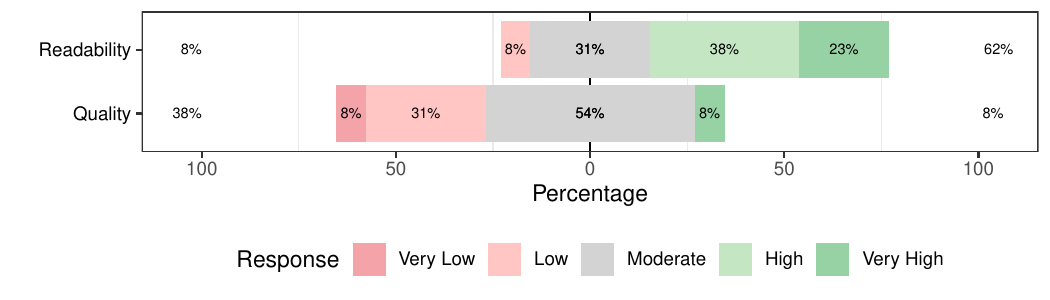}
		\caption{Diverging stacked barplot for responses of rates on readability and quality of generated tests using 5-points likert-scale collected in 2023. Note that positive responses (i.e., ``high'') extending to the right and negative responses (i.e., ``low'')  to the left from the diverging point.}
		\label{subfig:effEx2023Rates}
	\end{subfigure}\hfill
	\caption{Answers provided by industrial partners about readability and quality of tests generated by  \evo (QB1-2)}
	\label{fig:effRates}
\end{figure}

Regarding responses of \emph{QB}s, answers of \emph{QBs}1,2,5,7 and 8 are reported in Table~\ref{tab:effInfo} in detail for each participant.
Diverging stacked barplots for analyzing
rates on readability and quality are shown in Figure~\ref{fig:effRates}.
%
%

\qbA
Regarding the readability shown in Figure~\ref{subfig:effEx2021Rates}, we received 3 rates (60\%) on \textit{Low} and 2 rates (40\%) on \textit{High} in 2021.
All \textit{Low} rates are about \csA, and all \textit{High} rates are about \csB (see Table~\ref{tab:effInfo}).
Compared to the results collected in 2021, we achieve better rates on readability reported in Figure~\ref{subfig:appEx2023Rates}.
We received 1 rate (8\%) on \emph{Low}, 4 rates (31\%) on \emph{Moderate}, 5 rates (38\%) on \emph{High}, and 3 rates (23\%) on \emph{Very High}. 
Such improvements might be due to the native support of fuzzing RPC APIs.
In addition, suggested by our industrial partner, \evo \vsecond has been integrated with a simple test grouping strategy to put tests which represent faults (i.e., throwing exceptions) into a different file.
The strategy might help to have better rates on the readability. 
Regarding rates on different SUTs (see Table~\ref{tab:effInfo}),
the \emph{Low} rate is about \csD specified by a product manager. 
Note that the product manager provided a comment as ``it is unclear what functions the tests examine`` for answering \qbC, and it might be the reason for the \emph{Low} rate on readability.
For \csD, the other rates are 3 \emph{High} (by a development manager and two QA engineers) and 1 \emph{Very High} (by a QA engineer), then the rate results might also depend on their daily tasks, e.g., the product manager typically does not need to handle services at the source-code level, thus, they might have high requirements on the readability of the tests.
In addition, among the three APIs, we have the best rates of readability in \csC, i.e., 1 \emph{Moderate}, 2 \emph{High} and 2 \emph{Very High}.
For \csE, all of the three rates are \emph{Moderate}.   

\qbB
%
For \csA and \csB, in terms of overall quality, we received 4 rates (80\%) on \textit{Moderate} and 1 rates (20\%) on \textit{VeryHigh}.
In 2023, we received more negative rates (38\%) on the quality (i.e, 1 \emph{Very Low} and 4 \emph{Low}), and less positive rates (8\% with 1 \emph{Very High}).
This might be related to less code coverage we achieved on \csC, \csD and \csE, for instance, 3 negative rates specified on \csC (see Table~\ref{tab:effInfo}).
In addition, we found that all three participants who are in the development department (i.e., one development manager and two developers) gave negative rates.
Currently, \evo derives assertions based on responses.
The responses in the industrial APIs could be huge (hundreds or thousands of lines), then having too many assertions on each data entry might not help developers to easily identify the sources of the faults.
This might be a reason why we do receive negative rates from the developers.
How to improve \evo for better supporting developers will be a challenge we need to address.

We discuss \qbC and \qbD together as the two are correlated.
With the two questionnaires in 2021 and 2023, we received feedback on needs of further improvements on readability.
For instance, 
it would be better to have some endpoint or interface info on the names, which could help them to read the tests.
Grouping the tests according to interfaces and scenarios is highly requested, and it would be better to have further detailed clustering, such as under each interface, further clustering tests for input validation, business operation and exception thrown.
In addition, there exist many lines to instantiate input parameters of requests and assert responses.
To improve readability, it might be better to put such heavy info into a separated file, e.g., JSON.
Moreover, regarding assertions, one participant suggests that the assertions should be created specific to a scenario to test, and there is no need to assert every property of the responses.
As an automated approach, it is not trivial to decide which property should be shown, and this will require a further investigation in the future.

Other most common suggestions are related to how effectively cover their business logic.
Based on the feedback,  most tests generated by \evo properly cover the input validation, while only few tests cover their real business logic and important usage scenarios.

One of the advantages of manually written automated tests is easy-to-read,
e.g., test suites are often well grouped and do not contain many tests.
Another advantage is specific, i.e., 
the manually written automated tests are defined to address their business logic, and that  has more meaningful combinations of endpoints than automatically generated tests.
In addition, the tests could have assertions on database that is currently missing on the tests generated by \evo.

Furthermore, code coverage and efficiency to produce tests would be better to be improved. 
Besides, the suggestions on the generated tests, we also received one request to provide info about what code is not covered in terms of each interface, so that it would be easier for them to write manual tests for those missing cases.



\qbE and \qbF
Based on the results shown in 
Table~\ref{tab:effInfo}, with two experiments involving 18 answers, 16 participants showed
positive feedback in keeping the automatically generated tests, i.e., 100\% of the respondents in 2021, and 84\% of the respondents in 2023.
The 2 participants who responded ``No'' are from one developer who gave \emph{Very Low} rate on quality 
and one product manager who gave \emph{Very Low} rate on readability (see Table~\ref{tab:effInfo}).  

%
Regarding \textit{QB6},
the tests could be used together with manually written automated tests, and
they would like to further use the generated tests for robustness testing.
In addition, they also want to apply them for regression testing in their CI systems (i.e., each time when the services are updated).
The tests could also be used to validate and monitor their testing environment for ensuring that all services are up and running.


\begin{figure}
	\centering
	\begin{subfigure}{.47\textwidth}
		\vspace{-2\baselineskip}
		\includegraphics[width=.99\linewidth]{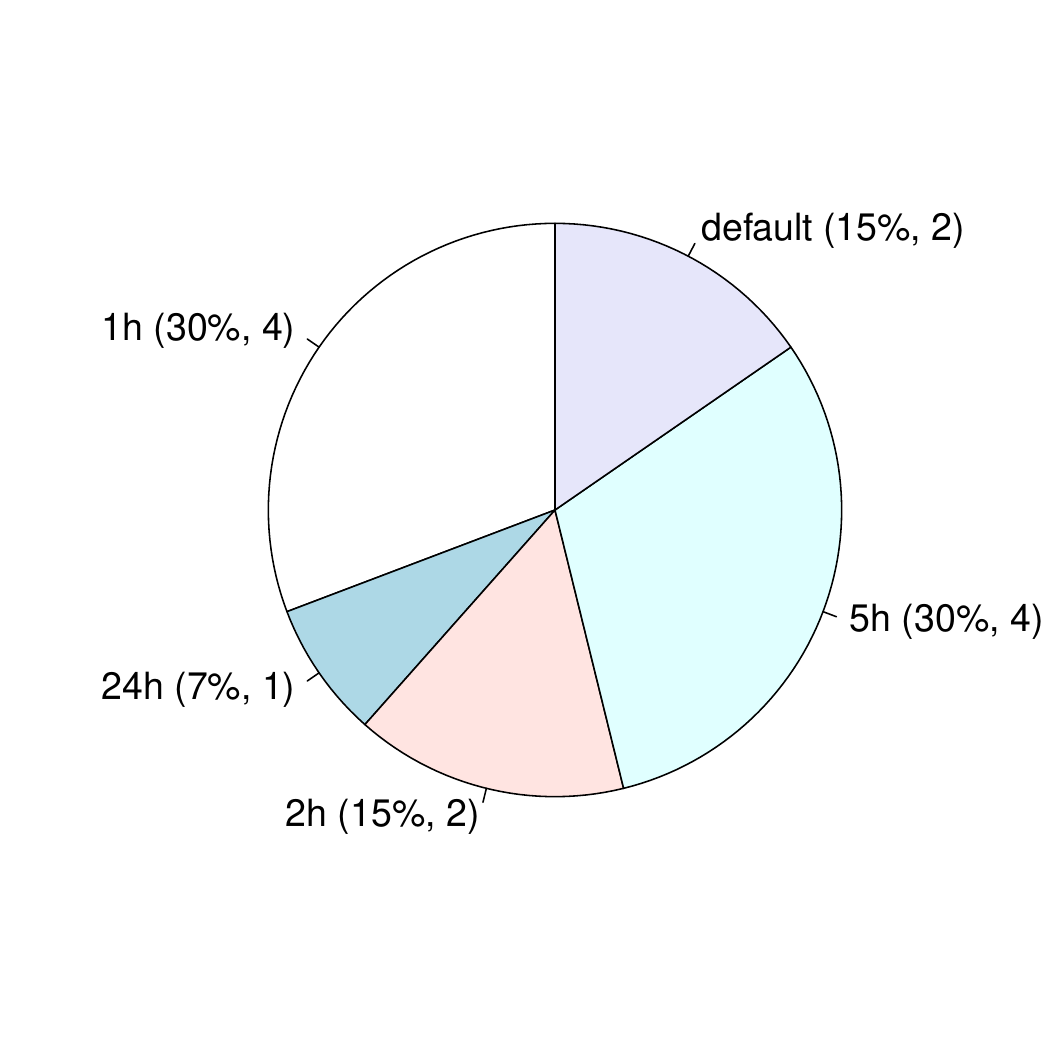}
		\vspace{-3\baselineskip}
		\caption{Results of preference of configuring time budget to use \evo by participants}
		\label{subfig:timePreference}
	\end{subfigure}\hfill
	\begin{subfigure}{.47\textwidth}
		\vspace{-2\baselineskip}
		\includegraphics[width=.99\linewidth]{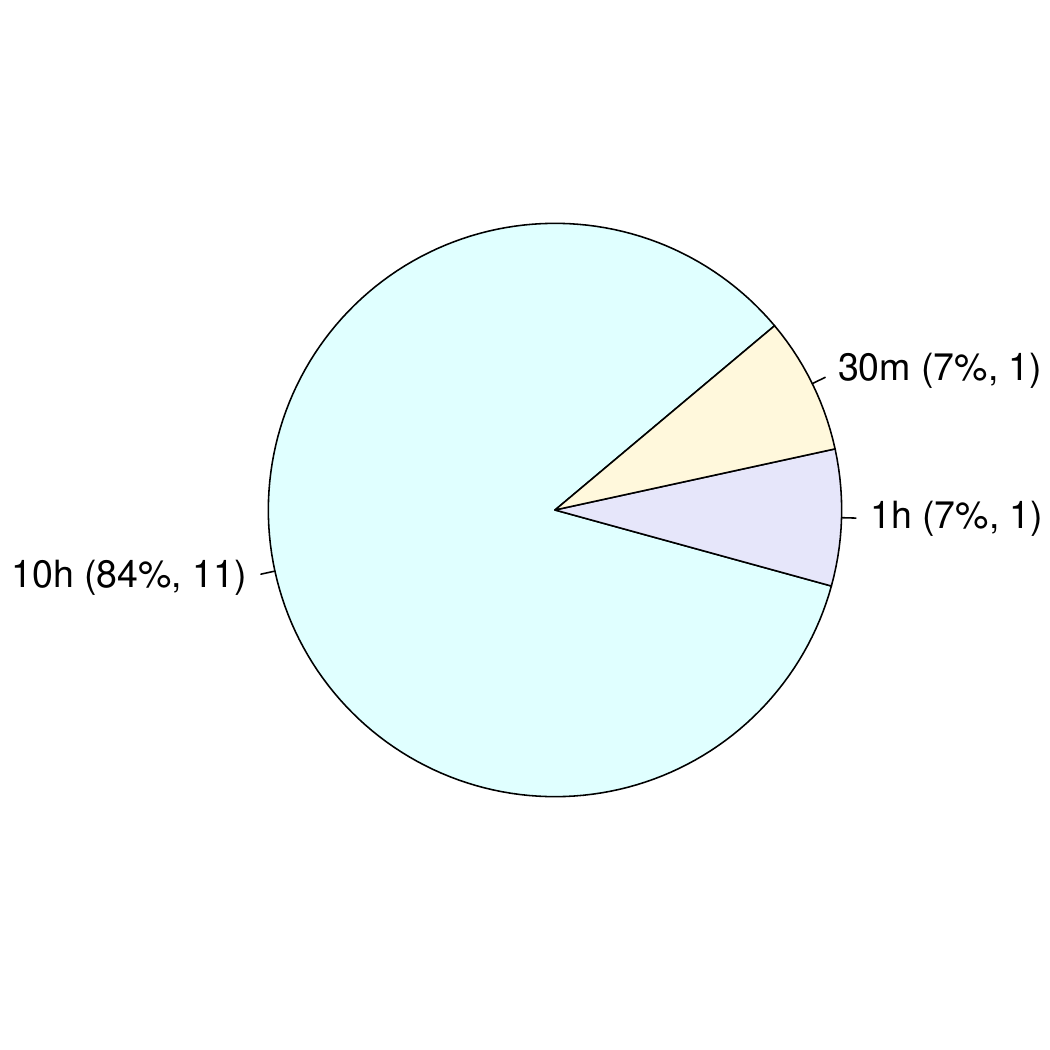}
		\vspace{-3\baselineskip}
		\caption{Results of selection of participants on the three time budget (i.e., 30m, 1h, 10h) options based on results}
		\label{subfig:effTimeOption}
	\end{subfigure}\hfill
	\caption{Results of feedback on time budget from industrial participants}
	\label{fig:timebudget}
\end{figure}

In the experiment with \evo \vsecond, we designed two further questions to collect responses related to time budget that practitioners would like to use.

\qbG
Based on results in Figure~\ref{subfig:timePreference}, 1-hour and 5-hours time budget are the most preferred options specified by the participants, i.e., 30\%.
There are 15\% participants who would like to use the \emph{default} setting and 2-hours time budget.
One participant responded 24 hours along with a comment as ``Unless the tool can achieve high coverage, there would not be problematic to apply the time budget such as more than 24 hours.''

\qbH
This question is designed to collect options of time budget by considering its performance (we provided results shown in Table~\ref{tab:rq2-coverage} in the questionnaire).
With the three time budget settings, 
as \evo with 10 hours could achieve better results, 84\% participants opted 10 hours, and 7\% participants opted 30 minutes and 1 hour. 

Based on the responses of \emph{QB7} and \emph{QB8}, from points of view of practitioners, they might concern more on quality of tests generated by \evo than time cost, as the tests are produced automatically.

\begin{result}
	
	\textbf{RO2.2}: Regarding results obtained at Meituan, with
	the recent version of \evo (\vsecond), most of the responses on readability are positive (i.e., 62\%) or neutral (31\%), while the rates on quality are 38\% negative, 54\% neutral and \%8 positive.
	There exist some limitations (i.e., business logic coverage, test readability and assertions on database state) that should be addressed for further improving the quality.
	Regarding options of time cost that a fuzzer (such as \evo) takes to generate tests, 1-hour and 5-hours are the most preference based on answers of 13 practitioners.
	 However, practitioners concern more on quality of generated tests compared to time cost (i.e., high time cost could be acceptable as long as it helps to improve the quality).
	 
	 	In addition, it is challenging to conduct such as a time-intensive user study on effectiveness of a tool at companies with which the researchers do not collaborate.
	 	All participants outside of Meituan claimed that they did not have enough time to inspect tests generated by \evo, thus, it is not feasible to generalize results we found at Meituan to them.
	 
\end{result}


\subsection{Results for RQ3: Integration}

Table~~\ref{tab:allques} (see \emph{QC}s) shows the questions we defined in our questionnaire and interview for inquiring potential integration and usage of \evo in industrial settings.
We reported feedback of major barriers in adopting \evo in User Study 2021 and User Study 2024 (Section~\ref{subsubsec:rq31}).
Since the first 
user study conducted in 2021, \evo has 
in 2023
been integrated into the testing pipelines at Meituan.
In Section~\ref{subsubsec:rq32}, we share the collaborative process of integrating 
\evo into industrial 
pipelines followed by a discussion about how 
practitioners
would like to use \evo based on responses collected in the three use studies in Section~\ref{subsubsec:rq33}.


\subsubsection{Results for RQ 3.1}
\label{subsubsec:rq31}

In the first 
user study at Meituan, we asked \emph{QC1} (\emph{\cqcA}) 
for collecting potential barriers of integration of \evo.
\#1--\#4 in Table~\ref{tab:majorChanges} presents four issues and problems answered by eight industrial participants.
The major barrier (\#1) was a lack of native support for RPC APIs used by Meituan. 
Without this support, automating driver generation becomes challenging as it requires manually creating an additional REST layer for RPC APIs. 
Moreover, using REST APIs for testing of RPC APIs caused side effects in fault detection, as the API calls were made via REST rather than RPC.
Due to the large volume of data in enterprise databases, it poses challenges in resetting enterprise APIs (\#2).
In addition, adopting a new technique into the development process may involve a learning curve for testers (\#3).
To use tests generated by \evo, employees at Meituan were concerned that the tests needed to be compatible with the enterprise's existing testing framework (\#4).

In User Study 2024, we received responses from three out of five participants, i.e., 1 DPM, 1 SWA, 1 Tester/QAM.
Two of them concerned the effectiveness of \evo in fuzzing their enterprise APIs.
Unfortunately, they did not identify extra benefits when applying \evo by themselves as discussed in Section~\ref{subsec:rq2}.
The remaining response pertained to learning curve as \#3 identified at Meituan.

\begin{result}
	\textbf{RQ 3.1}: We received responses regarding the learning curve as a major barrier to adopting \evo in both user studies.
	In User Study 2021 at Meituan, we also received three additional responses concerning the lack of native support for RPC APIs, the proper reset of enterprise APIs with large volumes of data in databases, and compatibility with the enterprise's testing framework. 
	The lack of native support for fuzzing RPC APIs was identified as the most important barrier.
	From outside of Meituan, the remaining responses related to the effectiveness of \evo, as they failed to identify extra benefits achieved by \evo.
\end{result}
	
\begin{table}
	\caption{Major updates our industrial partner and us made for enabling integration of \evo since 2021}
	\label{tab:majorChanges}
	\small
	\resizebox{1.0\linewidth}{!}{
		\begin{tabular}{p{0.01\linewidth} p{0.42\linewidth}  p{0.42\linewidth} p{0.1\linewidth}}
			\toprule
			\# & Problems  & Current status & Efforts \\
			\midrule
			1 & \emph{Since \evo does not have a native support for  Apache Thrift, this requires a manual configuration between Thrift and REST.
				This  would be a major adoption barrier from their point of view.} & This have been resolved. We have proposed a novel approach for supporting a white-box fuzzing of RPC APIs~\cite{zhang2023rpc}	& Researchers \\
			\midrule
			2 & \emph{Considering that there is no a viable solution to reset their databases yet, in order to generate independent tests with \evo, our industrial partner would need to develop such database reset operation from industrial side (as they use their own customized RDS).} & This is partially addressed. Our industrial partner developed a solution to give us access to execute all kinds of SQL commands that are for cleaning databases. But due to massive amount of data in databases and efficiency issues, we currently still cannot enable the reset in fuzzing web services. & Industrial partner\\
			\midrule
			3 & \emph{From the view of one of the principal software engineers/managers, there might exist a learning curve for their testers to adapt a new technique.} & Our industrial partners have enabled \evo into their pipeline, and there is not any manual efforts needed for running \evo on their industrial services. Thus, testers do not need to learn how to start and configure \evo.  & Industrial partner\\
			\midrule
			4 & \emph{From the view of one of the principal software engineers/managers, there is a need to extend \evo test case writer to adopt their testing framework.} & In~\cite{zhang2023rpc}, we defined a model specific to handling RPC interactions. Thus, based on tests specified with the model, our industrial partner is now extending \evo to produce JUnit tests which use their testing framework.
			However, as \evo produces self-contained JUnit tests (using \evo driver), the tests can still be executed on their pipeline, and they currently use the tests generated by \evo before the customized test writer is implemented. & Industrial partner\\
			\midrule
			5 & \emph{As the industrial APIs might employ some techniques developed in-house, \evo should provide some capability to industrial partners for further customization when needed.} & Since 2021, to better adopt to industrial settings, \evo has been extended to provide interfaces that allow users to customize methods to set up authentication, identify whether the SUT is up and running, implement own test case writer, and define a schema to categorize responses.   & Researchers\\
			\bottomrule
		\end{tabular}
		
	}
\end{table}
\subsubsection{Results for RQ 3.2}
\label{subsubsec:rq32}
	
To promote the adoption of an academic prototype (such as \evo) into the development pipeline of an enterprise, it requires efforts from both the enterprise and the researchers.

From the viewpoint of industry practitioners, they need to conduct an initial assessment of the tool, similarly to we did in the first user study (User Study 2021), before allocating more resources, such as employees and hardware.
In order to showcase potential benefits to our industrial partner, in User Study 2021,
we invited the industrial partner to assess our approach together, i.e.,
perform a manual review on tests generated by \evo with the best configuration (i.e.,10 hours) and coverage reports achieved by executing the tests.
With the review, the industrial partner found extra benefits achieved by
the tests outputted by \evo, as we discussed in Section~\ref{subsecsec:rq2s1}.
Such benefits gave our industrial partner confidence to take effort to use our approach and promote further its integration into their development/testing process.

As the authors of \evo, we addressed the most important challenge (i.e., \#1 in Table~\ref{tab:majorChanges}) by designing the first approach in the literature for enabling white-box fuzzing of modern RPC APIs
in 2022 (see received date of Reference~\cite{zhang2023rpc}).
In the process of integrating \evo, our
industrial partner also made great efforts to 
support the integration of
\evo into their testing pipelines, i.e., addressing the challenges of \#2--\#4 in Table~\ref{tab:majorChanges}.
After integrating \evo, we found 
another issue shown as \#5 in Table~\ref{tab:majorChanges}. As
the enterprise web services often use their in-house techniques, we have extended \evo in terms of customization for allowing users (such as our industrial partner) to use their own techniques, such as custom authentication setup and schema to identify whether a request has succeeded.
Note that such extension is meant to support customization for any users when needed.

To enhance the application of \evo at Meituan, in 2023, we conducted the second user study (i.e., User Study 2023) after \evo enabled a native support of fuzzing RPC APIs.
Compared to User Study 2021, we obtained more resources in User Study 2023.
For instance, experiment for assessing the effectiveness of \evo was conducted in a pipeline that allowed for 10 repetitions, rather than a single execution on an employee's machine.
More employees participated in the user study: 1 employee \emph{vs.}  10 for usability, 5 employees \emph{vs.} 15 for effectiveness, 8 employees \emph{vs.} 19 for integration and existing challenges.

\begin{result}
	\textbf{RQ 3.2}: Integrating an academic prototype into an industrial process requires collaborative efforts from both enterprises and researchers. 
	Industrial partners must conduct an initial assessment of the prototype before allocating resources, while researchers need to demonstrate the prototype's benefits to the industrial partners. 
	To promote an integration of \evo, we conducted User Study 2021 with employees at Meituan, and industrial partner identified extra benefits achieved by \evo. 
	Considering the benefits,
	our industrial partner is willing to integrate our tool in their development and testing processes. 
	All integration related barriers we identified in the first 
	user study have been resolved by our industrial partner and us.
	In 2023, \evo has been successfully integrated into the CI pipelines at Meituan, positively impacting hundreds of engineers and testers each day.
	To further enhance the application of \evo at Meituan, our industrial partner is willing to invest enterprise resources for conducting another user study.
\end{result}

\subsubsection{Results for RQ 3.3}
\label{subsubsec:rq33}
To further investigate how our industrial partner would like to use \evo, we asked
 \emph{QC2} (\emph{\cqcB} as shown in Table~\ref{tab:allques}) to all participants, across the three user studies.
 


In 
User Study 2021,  the preferred option for the industrial partner is to adopt \evo into their CI pipeline. 
As \evo has been integrated into their CI pipeline, 
in User Study 2023 the results were as follows: 
\begin{enumerate}[\itshape{AQC2}-1]
	\item Participants at Meituan would like to use \evo to generate tests automatically in their testing pipeline
	\begin{enumerate}
		\item when they create a pull request (QAM); 
		\item when code updates ; 
		\item when extra check is required to perform on services, such as before deployment into production, during smoke testing; 
		\item when there is a lack of manually written automated tests or low coverage is achieved by the tests (such as less than 20\%);
		\item as a scheduled task for testing services (additionally, such generated tests could be further used for validating and monitoring their testing environment, i.e., all services in the environment should be up and running).
	\end{enumerate}
	\item \label{aqc26} 
	In addition, two development managers and one developer also suggested to integrate \evo into their DevTools at Meituan;
	\item Moreover, three participants (development manager, QA manger and engineer) emphasized that they would like \evo to generated tests only for updated code;
	\item One participant (QA engineer) mentioned that she/he would like to replace manually written automated tests with tests generated by \evo.
\end{enumerate}

Compared to the responses obtained in 2021, we received more diverse responses in 2023.
This might relate to User Study 2023 involving participants from a wider range of positions and the CI integration of \evo.
	
In User Study 2024, we received a response from one out of five participants (i.e., SWA) expressing interest in integrating \evo into their DevTools for generating regression tests, similar to \emph{AQC2-\ref{aqc26}}.

\begin{result}
	\textbf{RQ 3.3}: Based on 27 responses 
	across two use studies at Meituan, as \evo has been integrated into their testing pipelines, they would like to use \evo in their daily tasks (such as Pull Request), trigger new test generation by \evo when needed (such as lack of manually written automated tests or low coverage achieved by the tests), or run \evo as a scheduled task to test systems in the microservices.
	In addition, industrial participants suggest to integrate \evo into their DevTools, have tests generated only for updated code, and replace manually written automated tests with tests generated by \evo.
	From outside of Meituan, we received a response from one out of five participants (i.e., SWA) who also expressed interest in integrating \evo into their DevTools as we found at Meituan.
\end{result}


\subsection{Results for RQ4: Existing problems and challenges}
\label{subsec:rq4}

The research task for RQ4 aims at understanding existing testing difficulties in industry, and discuss potential solutions that tools like \evo could help to tackle.
To achieve this goal, we defined three questions as shown in Table~\ref{tab:allques} and asked them to 
all industrial participants in the 
three user studies.


\qdA

Regarding responses from employees in the two user studies at Meituan, currently,
their testing tasks are mainly performed manually, e.g., manually write automated tests,  conduct manual testing by performing users' behaviors from client sides with UI, or analyze the SUT by replaying its historical execution.
All these manual tasks are time consuming, then they would like to have techniques to support automation for the following tasks.
\begin{enumerate}[\itshape{AQD1}-1]
	\item \label{aqd11} Based on the structure of their services and their business logic, they are willing to promote automation testing by considering at the level of APIs (such as multiple APIs), business logic, and UI (from Principal Software Engineer).
	\item \label{aqd12} Hundreds of the connected services are running on one platform, and the developers in a team typically do not know the details of services which they are not responsible for. Automated test data preparation is very appreciated as database and external services setup is very time-consuming for them (from all positions except product manger).
	\item They would like to have some kind of automated recommendations for possibly related services and database state for the SUT, which could help their testers (especially for new employees) to setup their testing scenarios (from Director, QA mangers and engineers).
	\item It also would be very helpful if there is an automated solution to provide info about what they misconfigure and what data is missing in the database (from developers, QA managers and engineers).
	\item \label{aqd15} As industrial requirements are often updated frequently (adding new features, or update existing features), they would like to have automated assessment on how updated requirements impact on current implementation (from development manager),  have generated tests according to requirements (from QA engineer), have generated tests according to updated code (from development manager, developer, QA manager and engineer), have automated solution to maintain tests due to frequent updates (from QA engineer).
	\item Since industrial APIs might have various critical features, they would like to have tests that can identify problems in terms of security, service characteristic and core services (from QA manager).
	\item They would like to have automatically generated tests which achieve more than 50\% line coverage (from QA manager and engineer).
\end{enumerate}

In User Study 2024, we received responses from three out of five participants outside of Meituan. 
Two of these participants (from QA manager and Tester/QA engineer) expressed willingness to automate test generation for API testing and business logic, which aligns with \emph{AQD1-\ref{aqd11}}.
One of these participants (from SWA) showed interest in automating the generation of unit tests and regression tests, as manually writing these tests is time-consuming, similar to \emph{AQD1-\ref{aqd15}}.

\qdB

Based on 
experience of participants,  
the hard to detect faults collected at Meituan in User Study 2021 and 2023 are related to:
\begin{enumerate}[\itshape{AQD2}-1]
	\item \label{aqd21} \textit{data in the database}: the services might perform some analysis on the existing data in the database, then results of the analysis could be used by other services. In this case, if there exist some dirty/invalid data in the database, it could crash the other processes.
	\item \label{aqd22} \textit{associated services}: in order to test the system regarding a business logic, there could be multiple services under test. An error might occur in one service (e.g., wrong data saved in database) but the fault might manifest only in other services (e.g., when reading such data, and expecting it to be correct). Considering the complexity of the whole system (hundreds of services), locating this kind of faults is hard.
	\item \textit{dependent frameworks}: to build their platform, they also used software frameworks, and these frameworks might have bugs which could result in errors of their services.
	\item \emph{non-functional fault, concurrent faults, and faults relating to extreme scenarios}: such faults could be detected during testing, but it is difficult to reproduce them before locating them.
	\item \emph{faults relating to non-core business}: it might not be easy to detect faults in implementation of non-core businesses as there typically exist rare tests for them.
	\item \emph{faults relating to complex computation}: they need to handle complex computation based on business needs. But it is impossible for them to test all possibilities as they are way too many. Thus, it would be difficult to find the faults existing in the possibilities which have not be tested.
	\sloppy
	\item \textit{non-deterministic faults}: since some faults occur non-deterministically, it is hard for them to reproduce them and fix them. This is a common issue in the testing of complex distributed systems.
	\item \label{aqd28} \textit{interaction with external environment}: Meituan's business has close interactions with humans (who are responsible for diverse tasks, e.g., quality inspector, storekeeper) and psychical devices referred as System Actors. The testing is performed typically with an assumption that the System Actors  operate properly. However, when it is not true,  then some faults would only be detected once the services are run in production.
\end{enumerate}

In User Study 2024, four out of five participants from outside Meituan stated that identifying faults associated with multiple services with respects to business logic and scenarios is challenging, consistent with \emph{AQD2-\ref{aqd22}}. 
The remaining participant pointed out that detecting faults related to exception handling in complex production environments can also be difficult, such as verifying whether data is properly rolled back in the event of a database exception, which might relate to \emph{AQD2-\ref{aqd21}} and \emph{AQD2-\ref{aqd28}}.

\qdC

In the two user studies, based on the responses to \emph{QC3}, there are six challenges that employees at Meituan face in testing:

\begin{enumerate}[\itshape{AQD3}-1]
	\item \label{aqc31} Based on the business 
	Meituan deals with, starting from a user request, there could go through lots of services that could result in very complex combinations of various inputs and different states of the system.
	Considering such complexity, it is hard to ensure quality of tests wrote by developers, and most  faults are identified by the QA department.
	\item In addition, it is hard to define the testing scope regarding such closely interacted 
	services.
	\item Some combinations of the services could be reached only under certain states of the system.
	To generate and explore such states is also challenging, e.g., constructing real scenarios would require a rich troves of data in the database.
	Currently, as mentioned by a principal software engineer who is responsible for developing the testing tools of Meituan, their team would need to construct data into their testing environment according to real scenario in production, e.g., based on real data from their services in production.
	Such data preparation in the database usually takes at least four hours each week.
	\item Moreover, their platform has a high demand for time-efficiency and security (e.g., payments) that would bring further challenges in testing.
	For instance, their database services are accessed by many services, and there could be slow SQL queries in one service that would negatively affect the database services, and so result in failures (e.g., timeouts) in other services.
	\item Furthermore, how to better test scenarios which is related to external environment is challenging for them, as discussed also in \textit{QD2}.
	Due to the rapid increase of business requirements, often there is limited time for performing manual  testing.
	Applied technology components and architecture could also be updated, and such updates might lead to conflicts and mismatches that bring further challenges in testing.
	\item Last but not least, there exist complex computations that cannot be tested sufficiently.
\end{enumerate}

Regarding \emph{QC3}, we received four responses from outside of Meituan. 
Three of the respondents indicated that the most important challenge is ensuring coverage of code, business logic, business scenarios, and requirements with their currently applied techniques (from 1 SWA and 2 Testers/QA engineers), associated with \emph{AQD3-\ref{aqc31}}.
One of the respondents also noted that balancing the testing budget is a challenge for them (from QA manager), however, such a challenge might heavily depend on the company.

\begin{result}
	{\bf RQ4:} Most  testing tasks are performed manually for testing the business logic, which shows an urgent need for effective automation support.
	Due to complexity of microservices and business features, there exist various challenges on, e.g.,
	identifying various faults,
	locating faults,
	defining test scope, handling external services and databases, preparing test data, treating frequent update of implementations with technique and architecture update, testing complex computation, testing under uncertain environment (such as humans), and balancing the testing budget.
\end{result}

\section{Lessons Learned and Challenges}
\label{sec:lessons}

\subsection{Testing setup}

In 
the two user studies conducted with our industrial partner, Meituan, the first challenge we faced was to set up our tool in the industrial environment (as discussed in Section~\ref{subsec:rq1}).
Unlike  open-source SUTs (like the ones collected in EMB~\cite{EMB,icst2023emb}), in industry when dealing with microservice architectures, relationships among services are more complex, accesses to databases are more restricted, and data in the database is more sensitive.
Thus, it is hard to handle all dependent resources and enable a proper reset of automated test generation in such industrial setting.

For external services, there could be a possibility to handle them with mocking techniques, and further manipulate responses with search from mocked services to maximize the coverage of testing targets (e.g., code coverage and fault detection).
However, considering the complexity of these dependencies, it would be challenging to setup them properly in an automated way.
Furthermore, this would further enlarge the search space for test case generation.
Note that these issues are not specific to just \evo, but they would likely apply to any fuzzer used in this testing domain (i.e., microservices).

Regarding the reset of databases, as discussed with the industrial partner, they could help us to isolate complete databases for testing with \evo.
As the first setup, they could provide empty databases with an interface to clean all data.
Thus, before executing every individual, we could reset the SUT to a certain state for generating independent tests at the end.
However, based on the answers collected during the survey and interviews, it might not be sufficient to generate tests starting with empty databases and, such tests cannot be executed in their main testing environment (i.e., cannot perform a reset of SUT in the main testing environment).
Considering the existing data in the database, they are intentionally added for covering their business scenarios in real practice.
Table~\ref{tab:sut-info} shows the number of rows of existing data in the databases the SUTs directly interact with in their testing environment.
There exist 256,024 rows for \csA, and 1,742,574 rows for \csE.
If the testing is involving many services, it is impractical to clean all data and re-add them before every single test.
To better adopt \evo into this kind of industrial setting, there would be a challenge about how to utilize such existing data in the fuzzer (e.g., during test sampling) that could further result in a good coverage on real scenarios.
This would be related to the data preparation challenge discussed in Section~\ref{subsec:rq4}, and have a proper reset of SUTs.



\subsection{Testing criteria}

By investigating code coverage reports and analyzing responses from industrial practitioners,
we found that it might not be sufficient to define testing targets only with code coverage, and fault detection, e.g., more code coverage might not result in covering more business scenarios.
For solving industrial testing problems, besides code coverage and fault detection, there would be a need to enhance the fitness function from more dimensions, e.g., business logic, time constraints, database performance and system security, and how to measure such dimensions and cope with them together is very challenging.

When analyzing the industrial services for the experiments in User Study 2021 at Meituan, 
we found that, in order to understand and monitor their system behaviors, the services are built with various
monitoring features. 
For instance, regarding the business logic, there exist a tracing infrastructure (named \textit{Tracer}) inspired by Google Dapper~\cite{sigelman2010dapper} that enables tracking a complete path taken through every services from a user request, as the example in Figure~\ref{fig:sut}.
With the code instrumentation done in \evo, such tracked paths could provide additional info for measuring coverage related to the business logic.
As a timely concern on database performance from our industry partners, it would be important to further enhance the fitness function with respects to SQL execution time for exploring slow SQLs in the SUT.



\subsection{Fault localization}

In microservices architecture, a request from the client  could go through a long and complex list of services.
The service where a fault is observed might not be the root of the fault, which could be in any interacted service with respect to this request.
Thus, locating the root of the fault is often challenging.
In the context of white-box testing, \evo manages to refer a last executed statement for an identified bug, i.e., 
exception thrown during execution and a server error identified based on responses customized by users if they exist~\cite{zhang2023rpc}, when testing a single web application.
However, to better locate faults in microservice architectures, more info would be required.
%
To help locating faults, it is required to have a more comprehensive study on faults in microservices with various aspects (e.g., security) and define more intelligent rules to identify the faults.
Moreover, non-deterministic faults often appear in microservice architectures, due to the nature of distributed systems.
Characterizing the causes of such non-deterministic faults (such as~\cite{Godefroid2019Flaky}) is required to be investigated in the context of microservice architectures.

\subsection{Assertion and readability}

Our current strategy to produce assertions in the generated tests is based on responses.
However, as testing of an industrial application, test assertions would be required to be more comprehensive.
As pointed out by our industry partner, we plan to firstly extend our assertions with respect to databases.
In addition, with this study, we found that some content in responses are non-deterministic.
A  proper handling is required for producing assertion to deal with such non-deterministic content, to avoid generating  flaky tests.

One major concern from testers/developers about automatically generated tests is readability.
A more readable output would help them to identify problems and locate faults.
\evo integrates a test clustering technique for splitting tests into different files with respect to faults in the context of REST APIs~\cite{marculescu2022faults}.
First, a further handling for RPC would be needed.
We now support splitting tests, which identified thrown exceptions, into a separated file.
Considering hundreds of tests (for example, for \csC), more clustering strategies would be required, e.g., splitting test suites grouped by interfaces, business scenarios.
Moreover, we could also simplify the generated tests by moving over-informative inputs and assertions into a separate file, as suggested by some of the testers at Meituan.
But whether such move would have side-effects on debugging will need to be investigated.
It would also be helpful to name tests with meaningful info, such as RPC interface name, or the scenario that the test examines.
Responses in industrial APIs could by very complex and huge, and not every properties are important to them. 
Assertions need to be optimized, but how to determine the optimization (such as based on ``importance'', but how to calculate ``importance''?) is required to study.
Furthermore, test readability can be one of the objectives to optimize for~\cite{daka2015modeling}.

\subsection{Generalization of Results}
\label{sub:generalizationside}

It is challenging to carry out such a time-consuming user study that requires to use the tool, run the experiment, and answer the questionnaire in companies where there is no formal collaboration agreement~\cite{ko2015practical}.
This is a common problem in software engineering research: \emph{``running user studies can be difficult, and researchers may lack solution strategies to overcome the barriers, so they may avoid user studies''}~\cite{davis2023s}.

One of the authors contacted few acquaintances in their professional network in China.
Five practitioners from different companies responded positively to participate in the third user study in 2024.
However, with this small sample size, it may not be sufficient to generalize our results obtained at Meituan to other companies.
Nonetheless, we discuss our findings by comparing responses from user studies conducted at Meituan and those conducted outside of Meituan in this section.


Regarding responses related to \emph{usability}, results at Meituan showed that the employees did not meet much difficulties in using \evo.
Although the writing drivers has the most \emph{Difficult} rates, the employees still managed to complete it. 
However, from outside of Meituan, only one participant out of five succeeded in writing a driver to enable white-box testing with \evo, and all participants highlighted the difficulty in configuring such a driver and stated the need for additional documentation and tutorials.

Regarding the most important feature to assess the \emph{effectiveness} of tests, code coverage received more votes than fault detection in the two user studies conducted at Meituan. 
In User Study 2024, which involved five practitioners from five different companies, code coverage also received slightly more votes than fault detection, similar to the observations at Meituan.
Regarding the important properties to rate \emph{readability} and \emph{quality} of tests, we observed the same winner in both user studies conducted at Meituan and outside of Meituan, i.e., grouping tests for readability, and coverage of business logic for quality.
Regarding responses to assessment of \emph{effectiveness}, 
as all five participants outside of Meituan stated that they did not have enough time to inspect the generated tests, it was not feasible for us to include their results and perform further analyses on them.

In terms of \emph{integration}, the learning curve seems to be one of major barriers concerned most by practitioners, as observed in both user studies conducted at Meituan and outside of Meituan.
In addition, we found that participants outside of Meituan concerned most about the effectiveness of \evo before making efforts to integrate it, while we did not observe such a concern in the user studies conducted at Meituan.
This might be because participants outside of Meituan failed to identify the benefits due to the complexity of conducting the experiment and the lack of approval to participate in the user study during their working hours. 
Benefiting from research-industry collaboration, all issues identified in adopting \evo have been resolved or partially resolved by our industrial partner and us.
Considering the lack of willingness to integrate \evo from participants outside of Meituan, we received only one answer from a participant who expressed interest to integrate \evo into their DevTools for automated regression test generation. 
This option was also identified in User Study 2023 conducted at Meituan.

Regarding answers relating to \emph{existing challenges} practitioners face, most of the challenges observed in User Study 2024 (i.e., outside of Meituan) have been reported in the user studies conducted at Meituan, except for one, i.e., balancing the testing budget.
However, such a challenge remains a fundamental challenge in software development and might also heavily depend on the company.

\subsection{Unit Testing Tool}
\label{sub:unit}
In this study, questionnaire we designed 
as shown in Table~\ref{tab:allques}
covers all questions from an industrial study on unit test generation~\cite{seip2017industrial}.
This was done to enable meta-analyses.
Note that, as none of participants in User Study 2024 fully completed all questions, we excluded their answers in these meta-analyses.

Based on results obtained in the two user studies from our industrial partner, compared
with the industrial feedback on the unit testing tools, both test automation tools (such as EvoSuite and \evo) obtain positive feedback on usability (\textit{QAs}), i.e., easy task to use it.
Regarding the generated tests (\textit{QB1}--\textit{QB5}), the unit testing tool received more negative answers on inputs and assertions than \evo.
This could happen because, in the context of system testing, combination of inputs for validation and responses dealt with are more complex than the unit testing.
Having more sufficient testing on input validation is also important for web services, and manually creating such combinations would be impractical.
Regarding assertions, 
in the first experiment, \evo utilizes the HTTP responses to create assertions, and assertions are created based on instantiated Java objects in tests for RPC in 2023.
The responses in web services are typically defined with certain purposes, e.g., 
info messages, while such meaningful responses are often not applied in unit testing.
In addition, both tools were evaluated as needing to have an improvement on readability of the generated tests, which is also related to the size of the generated test suite files.

Regarding \qbE, most of the answers for unit testing is \underline{No}, but we received 100\% \underline{Yes} answers from 5 participants in the first experiment and 84\% \underline{Yes} answers from 13 participants in the second experiment.
This result could be related to negative feedback on the generated tests by the unit testing tool.
In addition, compared with  system testing, for  unit testing it would be easier and less time-consuming to create unit tests manually, and that could result in the preference of developers/testers on manually written tests.
Regarding the integration (\textit{QC}), both surveys collected the same preferred option with CI, and the major barriers are both related to currently applied frameworks.
With \evo, these major barriers have been resolved by our industrial partner and us.

Due to the different complexity of the addressed problem compared to manual testing, the system testing automation tool (such as \evo) received a more positive feedback and shows a more promising  integration than unit testing generation. 
However, this is based on 
answers from employees at one company with which we have collaborated in industry, and more would be needed to be able to generalize this claim.


\subsection{Summary of Common Challenges}

In this section, we summarize important common challenges (denoted as C-\#) which Web API fuzzers might face in industrial settings, along with potential solutions and our future work.

\begin{enumerate}[{C}-1]
	\item To fuzz industrial Web APIs, besides generating inputs and requests to the SUT, the fuzzer requires to be capable of manipulating responses of the directly interacted services and databases.
	\begin{enumerate}[\futuredirection:]
		\item Mocking techniques are a suitable solution to handle external services.
		However, how to enable it as part of the fuzzer is a challenge that the research community needs to address.
	\end{enumerate}
	\item To enable an automated Web API testing, it requires to reset states of the SUT. However, enabling such reset is very challenging in industrial settings. For instance,
	databases in industry are very complex and large, e.g., more than 1 million rows for the test data in \csE. It is impractical to reset such database at every test execution.
	\begin{enumerate}[\futuredirection:]
		\item  As discussed with our industrial partner, they plan to implement a solution to flashback states of the databases based on timestamps, as they think such solution is also useful for them for their other testing tasks.
		However, the time efficiency of the flashback is of paramount importance, e.g., 1 second could result in at most 3600 requests within 1 hour, which would significantly limit the effectiveness of the fuzzer.
	\end{enumerate}
	\item In industry, testing data in the databases is typically maintained based on data collected in real production for testing their APIs with various business scenarios. Besides resetting databases,  considering the amount of data in these databases, how to make full use of them is another challenge.
	\begin{enumerate}[\futuredirection:]
		\item \evo is equipped with \textit{sql handling} to analyze existing data of SQL databases with JDBC connection.
		As a white-box SBST fuzzer, it is possible to enable heuristics for optimizing selection and manipulation of existing data which requests relate to.
		However, considering the large amount of data, how to optimize it in a time effective way would be a potential challenge we might face.
	\end{enumerate}
	\item 
	An industrial Web API could connect multiple databases using database sharding techniques, which is a common solution for distributed databases in industry.
	Thus, fuzzers are required to be able to handle multiple connections in their database handling solutions.
	\begin{enumerate}[\futuredirection:]
		\item
		\evo enables SQL database handling, but such handling is currently built based on one JDBC connection.
		\evo could be extended to support multiple JDBC connections, but how to cope with the multiple connections when inserting data (e.g., determine a database to perform the insertion) will be another challenge we need to address.
	\end{enumerate}
	\item  Common testing criteria might not be sufficient to tackle industrial problems.
	As discussed with our industrial partner, they concern more on metrics relating to their business logic, efficiency,  performance and security.
	Such dimensions are likely also common to other industrial APIs.
	How to measure these dimensions as testing criteria and deal with them during fuzzing APIs would be an important challenge that research community should address.
	\begin{enumerate}[\futuredirection:]
		\item  Currently \evo does not consider performance testing criteria, such as response time, which could be considered as future work to extend \evo for performance testing.
		However, such info has been collected during search.
		As requested by our industrial partner, \evo now can trace and output logs of all SQL commands executed during testing. Then, with this info, our industrial partner can conduct further performance analysis on their databases based on these logs.
		\item (specific to industrial microservices equipped with monitoring system)
		As a white-box fuzzer, it is possible to track paths of accessing APIs at runtime by instrumenting services for monitoring microservices, if the SUT has any such connections.
		Then, the tracked info might enable identifying business logic and further employ them as parts of fitness function of \evo.
	\end{enumerate}
	\item Due to close and complex interactions among Web APIs and databases, locating faults in microservices is a challenge industry currently meets.
	\begin{enumerate}[\futuredirection:]
		\item To better locate faults, it is the first step for researchers to understand and characterize faults in microservices, which is going to be an important future work.
	\end{enumerate}

	\item  Industrial Web APIs often employ sources of non-determinism in their implementation (such as calling \texttt{Random} or accessing current CPU clock) that can result in flaky tests.
	To avoid producing flaky tests, it is important that fuzzers can properly handle such sources of non-determinism in the generated tests.
	\begin{enumerate}[\futuredirection:]
		\item As a white-box fuzzer, \evo could take advantage of directly handling non-determinism in the source code (e.g., instrumenting a fixed seed for \texttt{Random}). However, it requires to first identify all possible non-deterministic sources in the source code.
	\end{enumerate}
	\item Hundreds of tests could be generated when fuzzing industrial APIs, such as 231 tests for \csA and 488 tests for \csC.
	Note: a fuzzer could evaluate millions of tests during its search, but the output is usually minimized, e.g., generate minimal test suites with the highest coverage achieved during the search.
	To better help testers use these generated tests (e.g., for debugging and regression testing), readability of the tests is important in industrial practice.
	\begin{enumerate}[\futuredirection:]
		\item Classification is one potential solution to improve readability of the tests, such as group tests based on success or failure responses into different files.
		The tests with failure in responses could be further classified into different files based on types (such as user error or system error).
		To enable such classification, a study for understanding faults in microservices is required to conduct first.
		\evo has been equipped with a clustering strategy for tests of REST APIs. For RPC, we developed a strategy to put tests which lead to exception thrown into a separated file. However, strategies specific to the domain of RPC APIs and business scenarios are needed.
	\end{enumerate}
\end{enumerate}

\section{Threats to Validity}
\label{sec:threats}


\textit{Conclusion Validity}. 
In this study,
we carried out three user studies
to assess \emph{usability} and \emph{effectiveness} of \evo, investigate \emph{potential integration} of \evo into industrial process, and understand \emph{existing challenges} from the viewpoints of practitioners.
These user studies were conducted with in total 32 industrial professionals 
with different roles, e.g., a director, QA mangers, testers, 
developers, developer managers and product manager.
To collect opinions and perceptions of practitioners,
we conducted questionnaire and interviews that comprise 29 questions,
and summarized our findings based on in total 606 answers of the questionnaire/interviews across the three user studies as shown in Table~\ref{tab:allques}.
As the authors of \evo, we 
also performed three tasks in this industrial evaluation, i.e., a preliminary study of \evo on industrial SUTs (such as database and authentication handling), coverage analysis based on generated tests, and share our collaborative process of integrating \evo into the industrial CI pipeline.
To reduce bias in the analysis of the results, we confirmed our findings with our industrial partner, ran the experiment of \evo on multiple enterprise APIs, and together with our industrial partner  performed a manual analysis in the generated tests for assessing the effectiveness of \evo.
In 2021, as \evo for experiments could only be run on the machine of an employee at Meituan, we conducted the first experiment with one single run.
However, as \evo has been integrated into the testing pipelines at Meituan in 2023, we conducted the second experiment with 10-time repetitions, to reduce issues with results obtained by chance due to the randomness nature of the search algorithm used in \evo.

\textit{Internal Validity}.
There is no guarantee that the implementation of \evo is bug-free. However, \evo has been used to perform many experiments using open-source SUTs~\cite{arcuri2019restful,arcuri2020sql,arcuri2021tt,zhang2021adaptive,zhang2023open,zhang2023javascript,zhang2023rpc}, and it is also carefully tested, i.e., currently with
735
test suites (including unit tests and
end-to-end tests), which covers
60.23\%
lines of its code base.
In addition, \evo is an open-source project which can be reviewed and examined by anyone who is interested in it.
Regarding the questionnaire survey, we have involved 32 industrial participants from 
eight profiles in industry 
(as discussed in Section~\ref{sec:design}) 
for reducing possible bias in these answers.

\textit{External Validity}.
In this work, we trade \emph{generability} (i.e., focus on one single company) and \emph{sample size} (e.g., limited number of engineers and testers, in total only 27) to obtain more \emph{realism}: i.e., we want to \emph{study the actual use of fuzzers in industry by practitioners on their own large, industrial systems}.
Alternatives would had been to use students in the empirical study, to have a larger sample size.
Not only there would be questions on how results could generalize to actual industrial practice, but there would be major barriers to setup such an experiment.
For example, how long time (possibly months) it would take for the students to get familiar with large industrial systems, part of a microservice with tens of millions of lines of code.
For the same reason, even if Meituan has hundreds of engineers and testers, given a specific API used as SUT, only a small number of these professionals are familiar with the details of such API.
Having an engineer from another department being involved in an experiment on an API that they never used / worked on before would not be realistic.
This is particular the case for when evaluating the \emph{readability} of the generated tests, as readability of the generated tests is strongly related as well on how familiar the tester is with the tested API.
In addition, for researchers, this is a typical problem, as it is hard to have an opportunity to transfer academic outputs to industry, and find more companies to conduct such type of empirical study.
Furthermore, this kind of analyses involving human subjects are time consuming, much more than for example evaluating a fuzzer only ``in the lab'' on a set of open-source APIs.
Also, there is a major difference between empirical studies with students and empirical studies with practitioners in industry, as their cost is much higher, and so usually their sample size is smaller.
However, this kind of studies are essential to address the gap between research and practice.
Many studies done only ``in the lab'' by researchers might rely on assumptions that do not hold in practice, making the use of novel scientific techniques not fit for practice.
A scientific result with no practical use might not be fitting for an engineering discipline like software engineering.
Reporting experience studies in different companies can help building a large enough body of knowledge throughout time, from which meta-analyses can be use to infer general results.
To this aim, we used the same type of questions from an existing work on the application of a unit test generation tool in industry (Section~\ref{sub:unit}).
In addition, the study was performed with Web APIs using the same RPC framework. 
In the context of RPC, different frameworks could share parts of common domain knowledge (e.g., database interactions and dependencies among services), and the findings in this study might be applicable in other contexts (such as Apache Dubbo).
However, five SUTs alone are not enough yet to draw general conclusions.
More user studies of this kind are in dire need in the literature.

\section{Conclusion}
\label{sec:conclude}

It is very challenging to test enterprise applications using a microservice architecture, as they are often composed of a large amount of web services, such as the ones developed at Meituan.
Automated techniques such evolutionary search have demonstrated their effectiveness on solving many testing problems. However, there is a lack of empirical evaluation in industrial microservices.

\evo is an open-source test case generation tool that exploits the latest advances in the field of Search-Based Software Testing for web services.
In this paper, we conducted an empirical study on integrating \evo into real industrial settings.
The study was performed 
three times, in 2021, 2023 and 2024, using three versions of \evo (i.e., \vfirst, \vsecond and \vthird) and involving 
in total 32
participants.
The first two user studies were carried out with 27 employees at Meituan, which is our industrial partner.
The third user study was carried out with 5 practitioners from five different companies that we have no formal collaboration agreements with.
The user studies were designed from viewpoints of practitioners that comprise questionnaire and interviews for assessing \emph{usability} and \emph{effectiveness}, \emph{integration} of \evo, and understanding \emph{existing challenges}  that practitioners are facing. 
Benefiting from research-industry collaboration, with an access to information of enterprise APIs and direct communication with their employees, we (as the authors of the academic prototype) discussed our findings and shared our experience in the process of integrating our tool into the industrial pipeline at Meituan.
%

Results show that \evo clearly demonstrates its benefits to our industrial partner, Meituan,
e.g., unused code identification, test generation, code coverage, and real fault detection.
But, there are still many critical challenges posed in this study that are required to be investigated by the research community, like 
interactions with external services and large distributed databases.

\evo is in active state of development. 
Since this pilot study at Meituan was carried out in 2021, we 
addressed the most major challenges for Meituan by proposing a novel approach in 2022 that facilitates
a native support for
RPC directly.
Once 
such major challenges 
pointed out in 
previous study 
are addressed, it 
is important to repeat these surveys/questionnaires with industrial practitioners, to see what next needs to be addressed before this kind of fuzzing techniques can become commonly used in industrial practice as the second user study we carried out at Meituan.

The third user study was performed for generalizing results obtained at Meituan to other enterprises.
Results on \emph{usability} and \emph{effectiveness} indicate the challenges in conducting the user study that requires programming proficiency to use the tool and takes time to run the experiment.
However, responses to \emph{integration} and \emph{existing challenges} can also be identified as in the user studies carried out at Meituan.

Many studies done only ``in the lab'' might rely on assumptions that do not hold in practice, making them unusable for industry.
This paper gives the important scientific contribution of empirically evaluating a Web API fuzzer in industry, confirming its applicability and usability in this domain.
Most of the challenges identified in this study are not specific to \evo, and would likely apply to any other fuzzer for Web APIs.

\evo is open-source, with each release automatically stored on Zenodo for long term storage (e.g., current version \vthird~\cite{zenodo200evomaster}).
To learn more about \evo, visit our website \emph{www.evomster.org}

\section*{Acknowledgments}
Man Zhang is supported by State Key Laborotary of  Complex \& Critical Software Environment (CCSE).
Andrea Arcuri is supported by the European Research Council (ERC) under the European Union’s Horizon 2020 research and innovation programme (EAST project, grant agreement No. 864972).


\bibliographystyle{ACM-Reference-Format} 

%

\bibliography{../../../papers}


\end{document}